\RequirePackage{ifpdf}
\documentclass[hyper,letterpaper]{JHEP3}
\usepackage{amsmath,amssymb,amsfonts,bm,amscd}
\usepackage{cite}
\usepackage{graphicx, wrapfig}
\usepackage{multirow}
\usepackage{verbatim}
\usepackage{appendix}
\usepackage{fancybox}
\usepackage{slashed}

\usepackage{url}
\usepackage{float}

\graphicspath {{figures/}}

\newcommand{\bea}{\begin{eqnarray}}
\newcommand{\eea}{\end{eqnarray}}
\newcommand{\be}{\begin{equation}}
\newcommand{\ee}{\end{equation}}

\newcommand{\Z}{{\mathbb Z}}
\newcommand{\R}{{\mathbb R}}
\newcommand{\C}{{\mathbb C}}

\newcommand{\Q}{{\mathbb Q}}

\newcommand{\Li}{{\rm Li}}

\def\Tr{{\rm Tr \,}}

\def\frak{\mathfrak}

\def\tilde{\widetilde}
\def\hat{\widehat}
\def\bar{\overline}

\def\CA{{\mathcal A}}
\def\CB{{\mathcal B}}

\def\CG{{\mathcal G}}
\def\CH{{\mathcal H}}

\def\CL{{\mathcal L}}
\def\CM{{\mathcal M}}
\def\CN{{\mathcal N}}
\def\CO{{\mathcal O}}

\def\CR{{\mathcal R}}

\def\CU{{\mathcal U}}
\def\CV{{\mathcal V}}
\def\CW{{\mathcal W}}

\renewcommand{\bar}{\overline}
\renewcommand{\hat}{\widehat}

\def\^{{\wedge}}
\def\*{{\star}}
\newcommand{\lsb}{\lfloor \!\!\!\!\: \lceil}
\newcommand{\rsb}{\rfloor \!\!\!\!\: \rceil}

\newcommand{\beq}{\begin{equation}}
\newcommand{\eeq}{\end{equation}}

\newcommand{\bbCP}{{\mathbb C \mathbb P}}
\newcommand{\T}{\mathbb{T}}
\def\CD{{\mathcal D}}

\def\CircleArrowleft{\ensuremath{
  \reflectbox{\rotatebox[origin=c]{180}{$\circlearrowleft$}}}}

\title{Equivariant Verlinde formula from fivebranes\\ and vortices}

\author{Sergei Gukov$^{1,2}$ and Du Pei$^{1}$\\
\\
$^1$ Walter Burke Institute for Theoretical Physics, California Institute of Technology, Pasadena, CA 91125 \\
$^2$ Max-Planck-Institut f\"ur Mathematik, Vivatsgasse 7, D-53111 Bonn, Germany}

\abstract{We study complex Chern-Simons theory on a Seifert manifold $M_3$ by embedding it into string theory. We show that complex Chern-Simons theory on $M_3$ is equivalent to a topologically twisted supersymmetric theory and its partition function can be naturally regularized by turning on a mass parameter. We find that the dimensional reduction of this theory to 2d gives the low energy dynamics of vortices in four-dimensional gauge theory, the fact apparently overlooked in the vortex literature. We also generalize the relations between 1) the Verlinde algebra, 2) quantum cohomology of the Grassmannian, 3) Chern-Simons theory on $\Sigma\times S^1$ and 4) index of a spin$^c$ Dirac operator on the moduli space of flat connections to a new set of relations between 1) the ``equivariant Verlinde algebra'' for a complex group, 2) the equivariant quantum K-theory of the vortex moduli space, 3) complex Chern-Simons theory on $\Sigma \times S^1$ and 4) the equivariant index of a spin$^c$ Dirac operator on the moduli space of Higgs bundles.
\\
\\
\\
\\
\\
\\
\\
\\
\\
\\
{\tt CALT-TH-2014-171}}

\begin{document}
\maketitle
\tableofcontents

\section{Introduction}

In recent years, there has been a lot of work on realizing conformal theories in two dimensions
and Chern-Simons theories with complex gauge groups on the world-volume of branes in string theory.
Most of these constructions, though, focus on ``non-compact'' (irrational) theories.
In particular, such a central element in two-dimensional CFT as the Verlinde formula \cite{Verlinde:1988sn}
has not yet found its home in supersymmetric brane configurations.

The Verlinde formula is a simple and elegant expression for the number of conformal blocks
in a 2d rational CFT on a Riemann surface $\Sigma$.
The number depends only on the topology of $\Sigma$, an integer number $k$ called the ``level'',
and a choice of a compact Lie group $G$ that in most of our discussion we assume to be simple.
For instance, when $\Sigma$ is a closed Riemann surface of genus $g$ and $G=SU(2)$ the Verlinde formula reads:
\be
\dim \CH (\Sigma_g;SU(2)_k) \; = \; \left( \frac{k+2}{2} \right)^{g-1} \; \sum_{j=1}^{k+1} \left( \sin \frac{\pi j}{k+2} \right)^{2-2g}.
\label{usualVerSU2}
\ee
This expression and its generalization to arbitrary $G$ have a number of remarkable properties.
First, for a fixed $g$, the expression on the right-hand side is actually a polynomial in $k$.
Moreover, even though the coefficients of this polynomial are, in general, rational numbers,
at every $k \in \Z$ it evaluates to a positive integer number (= number of conformal blocks).

The space $\CH$ that appears in the Verlinde formula \eqref{usualVerSU2} can be also viewed as the Hilbert space associated
to quantization of a symplectic manifold $(\CM_{\mathrm{flat}}(\Sigma;G), k\omega)$ that we briefly review in section~\ref{SUSYSide}.
Despite many realizations of quantization problems in superstring theory and SUSY field theories \cite{Gukov:2008ve,Dijkgraaf:2008fh,Nekrasov:2010ka,Gukov:2010sw,Yagi:2014toa},
a simple quantization problem that leads to \eqref{usualVerSU2} has not been realized.
In this paper, we not only realize the Verlinde formula \eqref{usualVerSU2} as a partition function of a certain brane system,
but we also propose its vast generalization based on the embedding in superstring theory.

In particular, we wish to re-create a ``complexification'' of the beautiful story that involves a number of exactly solvable theories, centered around the Verlinde formula:
\begin{subequations}\label{qYM}
\bea
\dim \CH (\Sigma;G,k) & = & Z_{\text{CS}} (S^1 \times \Sigma) \\
& = & \dim H^0 (\CM, \CL) \\
& = & \int_{\CM} e^{c_1 (\CL)} \wedge \text{Td} (\CM) \\
& = & Z_{G/G} (\Sigma) \\
& = & Z_{\text{A-model}} (\text{Gr} (N,k)) \\
& = & \dim \text{Hom} (\CB', \CB_{cc}) \\
& = & \dim \text{Hom} (\tilde \CB', \tilde \CB_{cc}).
\eea
\end{subequations}
The first line here simply follows from the fact that the problem of quantizing $(\CM_{\mathrm{flat}}(\Sigma;G), k\omega)$ is what one encounters in Chern-Simons gauge theory. The latter theory is topological \cite{Schwarz,Witten:1988hf}
and, therefore, has trivial Hamiltonian $H=0$, so that dimension of its Hilbert space
can be computed via path integral on $S^1 \times \Sigma$.
The second line is the result of geometric quantization of the moduli space $\CM = \CM_{\text{flat}} (\Sigma;G)$
of classical solutions with the prequantum line bundle $\CL$, and (\ref{qYM}c) follows from
a further application of the Grothendieck-Riemann-Roch theorem.

Then, (\ref{qYM}d) relates it to the partition function of the $G/G$ gauged WZW model \cite{Gerasimov:1993ws},
and (\ref{qYM}e) is based on the relation \cite{Witten:1993xi} to the partition function
(more precisely, a certain correlation function)
of the topological A-model on $\Sigma$ with the Grassmannian target space $\text{Gr}(N,k)$.
Finally, (\ref{qYM}f) and (\ref{qYM}g) follow from representing the Hilbert space $\CH^{\text{CS}}$
as the space of open strings in the $A$-model \cite{Gukov:2008ve} of complexification of $\CM$,
namely $\CM_{\text{flat}} (\Sigma; G_{\C})$, and in the $B$-model \cite{Gukov:2010sw} of its mirror $\CM_{\text{flat}} (\Sigma;{}^LG_{\C})$, where ${}^LG$ denotes the GNO or Langlands dual group.

Unlike the classical phase space $\CM = \CM_{\text{flat}} (\Sigma;G)$, its complexification $\CM_{\text{flat}} (\Sigma;G_{\C})$ is non-compact and, therefore, the corresponding Hilbert space $\CH (\Sigma;G_{\C},k)$ is infinite-dimensional. This fact is well known in the study of Chern-Simons theory with complex gauge group and all related problems where $\CM_{\text{flat}} (\Sigma;G_{\C})$ shows up. Thus, it is unclear what the analogue of \eqref{usualVerSU2} and \eqref{qYM} might be if we naively replace a compact group $G$ by its complexification $G_{\C}$. However, by identifying $\CM_{\text{flat}} (\Sigma;G_{\C})$ with the Hitchin moduli space, we argue that the infinite-dimensional Hilbert space $\CH (\Sigma;G_{\C},k)$ comes equipped with a natural $\Z$-grading:
\beq
\CH (\Sigma;G_{\C},k) \; = \; \bigoplus_{n\in\Z} \CH_n
\eeq
such that each graded piece, $\CH_n$, is finite-dimensional.
This allows us to introduce the graded dimension of $\CH (\Sigma;G_{\C},k)$, which we call the ``equivariant Verlinde formula'':
\beq
\dim_{\beta} \CH(\Sigma;G_\C,k) \; :=  \; \sum_n t^n \dim \CH_n,
\label{grdimH}
\eeq
where $t=e^{-\beta}$.
We then generalize each line in \eqref{qYM} and, in particular, formulate several new TQFTs in three and two dimensions that compute the graded dimension \eqref{grdimH}.
For example, for $G = SU(2)$, $g=2$ and large enough $k$, the equivariant Verlinde formula gives
\beq
\begin{split}
\dim_{\beta} \CH(\Sigma;G_\C,k) & =  \frac{1}{6}k^3+k^2+\frac{11}{6}k+1  \\
& + \left(\frac{1}{2}k^3+3k^2-\frac{1}{2}k-3 \right) t   \\
& + \left( k^3+8k^2-3k+6 \right) t^2  \\
& + \left(\frac{5}{3}k^3+16k^2-\frac{71}{3}k+6\right) t^3 \\
& + \left(\frac{5}{2}k^3+29k^2-\frac{109}{2}k+63\right) t^4 \\
& + \ldots,
\end{split}
\label{texpansionH}
\eeq
where a careful reader can recognize \eqref{usualVerSU2} as the degree-0 piece, {\it i.e.},
\beq
\CH_0 \; = \; \CH (\Sigma;G,k) \,.
\eeq
Also, one can verify that the coefficient of $t^n$ is always a positive integer, agreeing with its interpretation as dimension of $\CH_n$. (In writing the $t$-expansion \eqref{texpansionH} we assumed that $k$ is sufficiently large; the exact formula \eqref{EqVerSU2} is given in section~\ref{EVA} and always yields positive integer coefficients for all $k$.)

As we explain in the rest of the paper, the equivariant Verlinde formula provides a connection between SUSY theories --- {\it e.g.} realized on world-volume of various brane systems --- and quantization, namely quantization of compact spaces, such as $\CM_{\text{flat}} (\Sigma;G)$ and $\text{Bun}_G$, as well as their non-compact counterparts, such as $\CM_{\text{flat}} (\Sigma;G_{\C})$ and the Hitchin moduli space.
In particular, there are two 3d $\CN=2$ theories that will play an important role throughout this paper:
the so-called ``Lens space theory'' $T[L(k,1);\beta]$ and the mass deformation of a 3d $\CN=4$ sigma-model:
\beq
\begin{array}{c}
\text{3d}~\CN=2~\text{theory}~T[L(k,1);\beta] \\[.1cm]
\hline\hline
\text{super-Chern-Simons at level}~k \\
\text{with adjoint field}~\Phi~\text{of mass}~\beta
\\ \quad 
\end{array}
\qquad\qquad\qquad
\begin{array}{c}
\text{3d}~\CN=2~\text{theory}~T[\Sigma \times S^1; \beta] \\[.1cm]
\hline\hline
\text{sigma-model with target}~\CM_H \\
\text{and a real mass}~\beta~\text{for}~U(1)_{\beta} \\
\text{flavor symmetry}
\end{array}
\label{thetwothys}
\eeq
To compute the equivariant Verlinde formula, the first theory needs to be put on $\Sigma \times S^1$ and topologically twisted,
while the latter theory leads to an expression for \eqref{grdimH} in terms of the equivariant integral over the Hitchin moduli space.
The former is also equivalent to the IR limit of 3d $\CN=2$ SQCD with an adjoint multiplet
that can be found on the world-sheet of half-BPS vortex strings.
Thus, familiar vortex strings know about $t$-deformation of the Verlinde algebra!

Now we present a more detailed outline of the paper and summary of the results.

\subsection{Outline of the paper}

In section \ref{fivebranes} we state the problem and introduce a one-parameter deformation
of complex Chern-Simons theory on Seifert manifolds.

The two theories \eqref{thetwothys} are special cases of $T[M_3;\beta]$, where $M_3$ is an arbitrary Seifert manifold.
As we explain in section~\ref{fivebranes}, when $M_3$ is a Seifert manifold, the corresponding 3d $\CN=2$ theory $T[M_3]$
has a special flavor symmetry that we call $U(1)_{\beta}$.
Turning on the real mass $\beta$ for this flavor symmetry gives a family of 3d $\CN=2$ theories $T[M_3;\beta]$
which, via 3d-3d correspondence, provide a definition and natural regularization of complex Chern-Simons theory on $M_3$.
Then, in section~\ref{sec:twoapproaches}, we give the second, equivalent definition of complex Chern-Simons on $M_3$ as a standard topological twist
of the 3d $\CN=2$ theory $T[L(k,1);\beta]$ on $M_3$. (Evidence for this equivalence is presented in section~\ref{3dTQFT}.)

Section \ref{Brane} relates exactly soluble theories described in this paper to familiar brane constructions in type IIA and type IIB string theory.
On one hand, it will give us a concrete description of the Lens space theory $T[L(k,1);\beta]$ as summarized in \eqref{thetwothys} and,
on the other hand, will link our story to the classical problem about vortices on a plane, $\R^2 \cong \C$.
Non-compactness of the plane leads to non-compactness of the vortex moduli space,
which often is an obstacle in defining its topological and geometric invariants.
This problem is easily cured in the equivariant setting, equivariant with respect to the rotation symmetry of the plane.
In particular, it leads us to identify the equivariant quantum K-theory of the vortex moduli space with
the ``equivariant Verlinde algebra'' for complex Chern-Simons theory (whose explicit form is described in section~\ref{EVA})
and provides an analogue of (\ref{qYM}e).

Section~\ref{SUSYSide} gives a precise definition of the graded dimension \eqref{CSPartition} via 3d-3d correspondence
and shows that it can be written as an equivariant integral over the Hitchin moduli space.
This provides an analogue of (\ref{qYM}c).
The same graded dimension will be computed in other sections from a variety of different viewpoints.

In section \ref{3dTQFT} we demonstrate that $\beta$-deformed complex Chern-Simons theory is equivalent to a certain twist of 3d $\CN=2$ theory $T[L(k,1);\beta]$,
\beq
\boxed{{\text{twist of}~T[L(k,1);\beta]~\text{on} \atop \text{a Seifert manifold}~M_3}}
\quad = \quad
\boxed{{\text{$\beta$-deformed complex} \atop \text{Chern-Simons on}~M_3}}\, ,
\label{3dTQFTsummary}
\eeq
and compute its partition function \eqref{grdimviatwist} on $\Sigma \times S^1$ using the standard localization techniques. This gives a ``three-dimensional'' calculation of the equivariant Verlinde formula and, as such, can be regarded as a ``complexification'' of (\ref{qYM}a).

\begin{wrapfigure}{l}{0.5\textwidth}
\begin{center}
\includegraphics[width=0.35\textwidth]{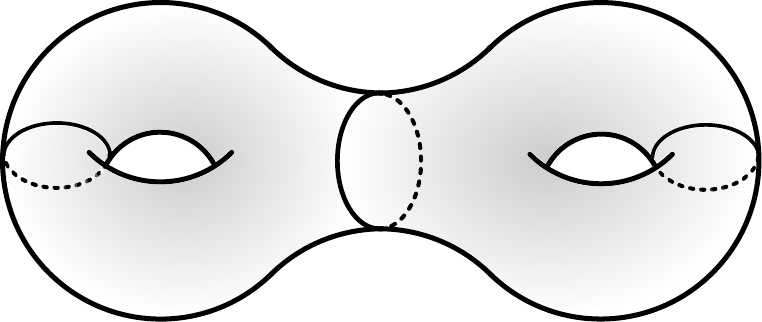}
\end{center}
\caption{A genus-2 Riemann surface can be decomposed into two pairs of pants.}
\label{fig:Genus2}
\end{wrapfigure}

The goal of section~\ref{TQFT} is to establish the analogue of (\ref{qYM}d) for the graded dimension \eqref{grdimH}. We call the resulting 2d TQFT the ``equivariant $G/G$ gauged WZW model''.
In section~\ref{EVA}, we formulate this theory as a set of gluing rules, by associating the ``equivariant Higgs vertex'' to each pair of pants, as in figure~\ref{fig:Genus2}.
In this section, we also discuss $t$-deformation and categorification of the Verlinde algebra.

In fact, 3d and 2d topological theories that compute \eqref{grdimH} are part of a larger family of TQFTs labeled by $R \in \Z$.
In three dimensions, $R$ can be identified with the R-charge of the adjoint multiplet $\Phi$ in the twisted theory $T[L(k,1);\beta]$.
This leads to a generalization of \eqref{3dTQFTsummary}.
Via reduction to two dimensions, we obtain a large family of new TQFTs that generalize the gauged WZW model.
Certain special values of $R$ correspond to models that have been previously studied from different viewpoints.

{}From this perspective, sections~\ref{SUSYSide} and \ref{EVA} are all about the special case $R=2$.
Section~\ref{3dTQFT} talks about general $R$, but most of the concrete formulae are written for $R=2$.
This is well compensated in section~\ref{TQFT}, whose main goal is to describe the family of 2d TQFTs on $\Sigma$ for general $R$.
Then, in section~\ref{sec:Rtwo} we again focus on $R=2$ that gives the equivariant $G/G$ model (EGWZW)
and whose partition function computes the equivariant Verlinde formula.
Similarly, in section~\ref{sec:Rzero} we focus on $R=0$ which gives the $G/G$ gauged WZW-matter model (GWZWM).

In total, in this paper we present {\it five} independent and concrete ways to compute the equivariant Verlinde formula:

\begin{itemize}

\item
a three-dimensional computation in a topologically twisted 3d $\CN=2$ theory on $M_3 = \Sigma \times S^1$ (section~\ref{3dTQFT});

\item
a computation based on 3d-3d correspondence that leads to an equivariant integral over the Hitchin moduli space $\CM_H$ (sections~\ref{SUSYSide} and \ref{EVA});

\item
a two-dimensional computation in the equivariant $G/G$ model on $\Sigma$ (section~\ref{sec:Rtwo});

\item
another two-dimensional calculation in the abelian 2d theory on the Coulomb branch (section~\ref{sec:BetheGauge});

\item
yet another two-dimensional calculation based on pair-of-pants decomposition of $\Sigma$
and the ``equivariant Higgs vertex'' (section~\ref{EVA}).

\end{itemize}

\noindent
It would be nice to add to this list a computation based on the 4d $\CN=2$ Lens space index \cite{Alday:2013rs,Razamat:2013jxa}.
Also, in section~\ref{sec:vortices} we outline a generalization of (\ref{qYM}e) that allows us to compute the equivariant Verlinde formula in the twisted theory on the vortex world-sheet. It would be nice to carry out the details of this approach and make contact with the equivariant vortex counting in \cite{DGH}.


\section{Fivebranes on Riemann surfaces and 3-manifolds}
\label{fivebranes}

Our starting point is the following configuration of M-theory fivebranes:
\beq
\begin{matrix}
{\mbox{\rm space-time:}} & \qquad & L(k,1)_b & \times & T^* M_3 & \times & \R^2\\
& \qquad &  &  & \cup \\
N~{\mbox{\rm fivebranes:}} & \qquad & L(k,1)_b & \times & M_3\, ,
\end{matrix}
\label{3d3d1}
\eeq
that is also used {\it e.g.}~in 3d-3d correspondence. Here, $M_3$ is an arbitrary 3-manifold, embedded in a local Calabi-Yau 3-fold $T^*M_3$ as the zero section. As a result \cite{Bershadsky:1995qy}, the three-dimensional part of the fivebrane world-volume theory is topologically twisted. Namely, the topological twist along $M_3$ is the so-called Blau-Thompson twist \cite{Blau:1996bx,Blau:1997pp}. It preserves four real supercharges on the fivebrane world-volume, so that the effective theory in the remaining three dimensions of the fivebrane world-volume (which are not part of $M_3$) is 3d $\CN=2$ theory. This theory is usually denoted $T_N [M_3]$ since it depends on the number of fivebranes in \eqref{3d3d1} and on the choice of the 3-manifold $M_3$. (Sometimes, one simply writes $T[M_3]$ when the number of fivebranes is clear from the context, or denotes this theory $T[M_3;G]$.)

The effective 3d $\CN=2$ theory $T_N [M_3]$ can be further put on a curved background \cite{Festuccia:2011ws,Imamura:2011wg}, in particular on a squashed Lens space $L(k,1)_b$:
\beq
L(k,1)_b \; := \; \{ (z,w) \in \C^2 \,, ~b^2 |z|^2 + b^{-2} |w|^2 = 1 \} / \Z_k \,,
\eeq
where the action of $\Z_k$ is generated by $(z,w) \mapsto (e^{2\pi i /k} z , e^{-2\pi i /k} w)$.
Then, reversing the order of compactification,
it has been shown \cite{Cordova:2013cea,Lee:2013ida} that the effective 3d theory on $M_3$
is the complex Chern-Simons theory,
confirming the conjecture of \cite{DGH,Dimofte:2011ju} (see also \cite{Terashima:2011qi,Cecotti:2011iy,Dimofte:2011py,Yagi:2013fda,Dimofte:2014zga}).

Therefore, one can reduce the six-dimensional $(2,0)$ theory in two different ways, summarized by the following diagram:
\beq
\begin{array}{ccccc}
\; & \;\text{6d $(2,0)$ theory} &\text{on}  &\text{$L(k,1)_b\times M_3$}  \; & \; \\
\; & \swarrow \qquad\qquad\qquad& \; &\qquad\qquad\qquad \searrow & \; \\
\text{3d $\CN=2$ theory $T[M_3]$} & \; & \; & \; & \text{complex Chern-Simons} \\
\text{on $L(k,1)_b$} & \; & \; & \; & \text{ theory on $M_3$}
\end{array}
\label{3d3d0}
\eeq
The statement of 3d-3d correspondence is that physics of complex Chern-Simons theory on $M_3$ is encoded
in the protected (supersymmetric) sector of the 3d $\CN=2$ theory $T [M_3]$.
For instance, SUSY vacua of the theory $T[M_3]$ are in one-to-one correspondence with the complex flat connections on $M_3$.
Various supersymmetric partition functions of $T[M_3;G]$ compute quantum $G_{\C}$ invariants of $M_3$,
{\it e.g.}~the vortex partition function (on $\R^2_{\hbar} \times S^1$)
gives the perturbative partition function of complex Chern-Simons theory labeled by a flat connection $\alpha$:
\beq
Z_{T_N[M_3]}^{\mathrm{vortex}}(\hbar,\alpha) \; = \; Z_{CS}^\alpha(M_3;\hbar) \,.
\eeq
Similarly, and closer to the setup in \eqref{3d3d1} that we shall use in this paper, the partition function of 3d $\CN=2$ theory $T[M_3]$
on the squashed Lens space is equal to the full partition function of complex Chern-Simons theory on $M_3$ at level $(k,\sigma=k\frac{1-b^2}{1+b^2})$:
\beq
Z_{T_N[M_3]}\left[L(k,1)_b\right] \; = \; Z^{(k,\sigma)}_{\mathrm{CS}} \left[M_3;GL(N,\C)\right] \,.
\eeq
This correspondence, relating partition functions of a supersymmetric theory with those of a TQFT, is obviously a very interesting one. However, there is much to be understood on both sides. On the right-hand side, one basic problem is to produce a simple and effective technique to compute the partition function of complex Chern-Simons theory on arbitrary 3-manifolds (see \cite{Dimofte:2009yn,Dimofte:2011gm,Gukov:2011qp} for some steps in this direction). On the ``supersymmetric'' left-hand side of the 3d-3d correspondence, the main problem is to develop tools for building the theory $T_N[M_3]$ associated with a given $M_3$. Previous attempts to tackle this problem either address only a certain sector of the theory $T_N[M_3]$ that does not capture all SUSY vacua / flat connections \cite{Dimofte:2011ju,Dimofte:2011py} or build the full theory $T_N[M_3]$ only for particular 3-manifolds \cite{Chung:2014qpa} and, therefore, are not systematic.

In particular, one motivation for our work is to understand $T_N [M_3]$ for Seifert 3-manifolds which, aside from the abelian case discussed in \cite[sec. 2.2]{Gadde:2013sca}, escaped attention in 3d-3d correspondence. A Seifert manifold is the total space of a circle V-bundle over a two-dimensional, closed and orientable orbifold $\Sigma$,
\beq
\label{Seifert}
S^1 \lhook\joinrel\longrightarrow M
\buildrel\pi\over\longrightarrow\Sigma\,.
\eeq
Although all computations in this paper can be easily generalized to arbitrary Seifert manifolds, for simplicity and concreteness we often carry out explicit computations in the basic example of a product $M_3=\Sigma\times S^1$ explaining how generalizations can be achieved.

With $M_3=\Sigma \times S^1$, the eleven-dimensional geometry \eqref{3d3d1} becomes:
\beq
\begin{matrix}
{\mbox{\rm symmetries:}} & \qquad &   &   & \!\!\!\! U(1)_N &   &   &   & SU(2)_R\\
& \qquad &   &  &\!\!\!\!\!\!\! \CircleArrowleft &  &  &  &  \CircleArrowleft \\
{\mbox{\rm space-time:}} & \qquad & L(k,1)_b & \times & T^* \Sigma & \times & S^1 & \times & \R^3 \\
& \qquad &   &  & \cup \\
{\mbox{\rm $N$ fivebranes:}} & \qquad & L(k,1)_b & \times & \Sigma & \times &S^1
\end{matrix}
\label{3d3d2}
\eeq
Now, one needs to do the topological twist only along a Riemann surface $\Sigma$ which is embedded in the local Calabi-Yau 2-fold $T^*\Sigma$ in a supersymmetric way as the zero section. In particular, it preserves half of supersymmetry on the fivebrane world-volume, which now also includes the $S^1$ factor. Interpreting this $S^1$ as the M-theory circle, the above system of fivebranes reduces to $N$ D4-branes, which carry maximally supersymmetric 5d super-Yang-Mills on their world-volume. A further reduction of 5d super-Yang-Mills on a Riemann surface with a partial topological twist along $\Sigma \subset T^* \Sigma$ requires gauge field and its superpartners to obey certain equations on $\Sigma$. This partial twist was studied exactly 20 years ago \cite{Harvey:1995tg,Bershadsky:1995vm} and the corresponding BPS equations turn out to be the Hitchin equations \cite{Hitchin:1986vp}, so that the effective 3d $\CN=4$ theory is a sigma-model with Hitchin moduli space $\CM_H(\Sigma;G)$ as the target. In recent years, this setup was also used in connection with the geometric Langlands program, AGT correspondence, {\it etc.}

To summarize, when $M_3=\Sigma \times S^1$, the effective 3d theory $T_N[\Sigma\times S^1]$ has $\CN=4$ supersymmetry and the R-symmetry group is enhanced to $SU(2)_R\times SU(2)_N$. A subgroup of this R-symmetry group can be easily identified with isometries of the M-theory geometry: $SU(2)_R$ is the double cover of the rotation group $SO(3)$ acting on the last factor $\R^3$ in (\ref{3d3d2}), while $U(1)_N$ (= a subgroup of $SU(2)_N$) acts on the cotangent fiber of $T^*\Sigma$.

One can introduce new parameters by weakly gauging these symmetries.
We will be interested in a ``canonical mass deformation'' of $T[\Sigma\times S^1]$ which gives a $\CN=2$ theory that in \eqref{thetwothys} we denoted $T[\Sigma\times S^1;\beta]$. This deformation can be done to any 3d $\CN=4$ theory by regarding it as a 3d $\CN=2$ theory, whose R-symmetry group $U(1)_{R'}$ is generated by $j^3_N+j^3_R$, and weakly gauging $U(1)_\beta$ generated by $j^3_N-j^3_R$.
Here we use $j^i_{N,R}, i=1,2,3$ to denote the generators of $SU(2)_N\times SU(2)_R$.

Note, from the viewpoint of $\CN=2$ supersymmetry, $U(1)_\beta$ is a flavor symmetry that acts on the sigma-model target $\CM_H(\Sigma;G)$ as
\beq
U(1)_{\beta} : \quad (A,\Phi) \mapsto (A, e^{i \theta} \Phi),
\label{Hitbsymm}
\eeq
where each point in $\CM_H(\Sigma;G)$ is represented by a Higgs bundle $(A,\Phi)$, see section~\ref{SUSYSide} for a brief review.
Weakly gauging this $U(1)_{\beta}$ symmetry deforms $\CN=4$ sigma-model with target $\CM_H(\Sigma;G)$ to a $\CN=2$ theory $T[\Sigma \times S^1; \beta]$ with the same field content, but where half of the fields have (real) mass $\beta$.
This deformation of $T[\Sigma\times S^1]$ can be realized in the brane geometry (\ref{3d3d2}) by introducing $\Omega$-background on both the two-dimensional cotangent fiber of $T^*\Sigma$ and on $\R^2 \subset \R^3$ with the equivariant parameters $\beta$ and $-\beta$, respectively.
We continue the discussion of the 3d $\CN=2$ theory $T[\Sigma \times S^1; \beta]$ in section~\ref{SUSYSide}.

Now, let us consider what this deformation means on the other side of the 3d-3d correspondence, {\it i.e.}~for the complex Chern-Simons theory on $M_3$. When $M_3=\Sigma \times S^1$ and $\beta=0$ we have ordinary (undeformed) complex Chern-Simons theory, whose partition function on $\Sigma\times S^1$ computes the dimension of the Hilbert space associated to $\Sigma$:
\beq
Z_{\mathrm{CS}}[\Sigma\times S^1;G_\C] \; = \; \mathrm{dim}\,\CH_{\mathrm{CS}}(\Sigma;G_\C) \,.
\label{Zdimxx}
\eeq
The space $\CH_{\mathrm{CS}}(\Sigma;G_\C)$ is infinite-dimensional due to non-compactness of the gauge group and one needs to regularize it in order to make sense of the above expression. We will do so by considering the graded dimension with respect to a $\Z$-grading on $\CH_{\mathrm{CS}}(\Sigma;G_\C)$ induced by the circle action $U(1)_{\beta}$. We call the resulting TQFT the ``$\beta$-deformed complex Chern-Simons theory''. Note that the $\beta$-deformed complex Chern-Simons theory is well-defined not only on $\Sigma\times S^1$ but also on arbitrary Seifert manifolds since this is the class of 3-manifolds for which one finds the extra symmetry $U(1)_\beta$.

In order to understand how $U(1)_\beta$ acts on the fields of complex Chern-Simons theory, we can follow {\it e.g.}~\cite{Cordova:2013cea} and reduce the six-dimensional $(2,0)$ theory on the Hopf fiber of $L(k,1)$ to obtain 5d $\CN=2$ super-Yang-Mills theory on $S^2 \times (\Sigma\times S^1)$. The Lorentz and R-symmetry group $SO(5)_L\times SO(5)_R$ of the five-dimensional theory is broken down to
\beq
SO(5)_L\times SO(5)_R \; \to \; SO(2)_L\times SO(3)_L \times U(1)_N \times SU(2)_R \,.
\eeq
Here $SO(2)_L$ is the Lorentz symmetry factor associated with $S^2$, while the second $SO(3)_L$ is the Lorentz factor associated with $\Sigma \times S^1$. If we choose the metric on $\Sigma$ to be independent of $S^1$, the holonomy group is reduced from $SO(3)$ to $U(1)$. So in order to do the topological twist along $\Sigma$, we only need to use a $U(1)_L$ subgroup of $SO(3)_L$ and identify the new Lorentz group $U(1)'$ with the diagonal subgroup of $U(1)_L\times U(1)_N$. Also, the $\Omega$-background picks out a $U(1)_R$ subgroup of $SU(2)_R$. In table~\ref{Fields}, we summarize how the fields in 5d super-Yang-Mills decompose and transform under $SO(2)_L\times U(1)_L \times U(1)_N \times U(1)_R$.

\begin{table}
\centering
\begin{tabular}{|c|c|c|c ccc c cc|}
\hline
5d & $SO(5)_L\times SO(5)_R$ & field &  $SO(2)_L$ &$\times$& $U(1)_L$ & $\times$ & $U(1)_N$ & $\times$ & $U(1)_R$\\
\hline
\hline
 & & $A$ & 0 & & $\pm 2$ & & 0 & & 0 \\
\cline{3-10}
$A^{5d}$ & $(5,1)$ & $A_0$ & 0 & & 0 & & 0 & & 0 \\
\cline{3-10}
 & & $B$ & $\pm 2$ & & 0 & & 0 & & 0\\
\hline
& & $\phi$ & 0 & & 0 & & $\pm 2$ & & 0 \\
\cline{3-10}
$\phi^{5d}$ & $(1,5)$ & $\phi_0$ & 0 & & 0 & & 0 & & 0\\
\cline{3-10}
& & $Y$ & 0 & & 0 & &  0 & & $\pm 2$ \\
\hline
$\lambda^{5d}$ & $(4,4)$ & $\lambda$  & $\pm1$ & & $\pm1$ & & $\pm1$ & & $\pm1$\\
\hline
\end{tabular}
\caption{The spectrum of 5d $\CN=2$ super-Yang-Mills theory on $S^2\times\Sigma\times S^1$.}\label{Fields}
\end{table}

After the topological twist, the scalar $\phi$ becomes a one-form on $\Sigma$. In fact, $\CA=A+i\phi$ and $\CA_0=A_0+i\phi_0$ can be identified with the components of the connection of complex Chern-Simons theory along the $\Sigma$ and $S^1$ directions, respectively. The $U(1)_\beta$ symmetry \eqref{Hitbsymm} does not act on $A$, $A_0$ or $\phi_0$ but acts on $\phi$ by rotating its two components $(\phi_1,\phi_2)$:
\beq
\label{Uonethetab}
\theta \in U(1)_{\beta} ~:\quad \left(\begin{array}{c}
\phi_1\\
\phi_2
\end{array}\right) \mapsto \left(\begin{array}{c}
\cos\theta\cdot\phi_1-\sin\theta\cdot\phi_2\\
\sin\theta\cdot\phi_1+ \cos\theta\cdot\phi_2
\end{array}\right).
\eeq
As it is precisely $\phi$, the imaginary part of the complex gauge connection, that gives rise to divergence in \eqref{Zdimxx},
one might hope that the $\Z$-grading of the Hilbert space $\CH_{\mathrm{CS}}(\Sigma;G_\C)$ induced by $U(1)_\beta$ symmetry could provide
the desired regularization. Indeed, as we show below, for each value of the $\Z$-grading, the corresponding component of the Hilbert space $\CH_{\mathrm{CS}}(\Sigma;G_\C)$ is finite-dimensional, so that the partition function of the $\beta$-deformed complex Chern-Simons theory is a polynomial in $t=e^{-\beta}$ that gives the graded dimension \eqref{grdimH} of the Hilbert space:
\beq
\label{CSPartition}
\mathrm{dim}_\beta\,\CH_{\mathrm{CS}}(\Sigma;G_\C) \;=\; Z_{\mathrm{CS}}[\Sigma\times S^1;G_\C, \beta] \,.
\eeq
The coefficient of $t^n$ counts the dimension of the subspace that has eigenvalue $n$ with respect to the symmetry $U(1)_\beta$.

In Chern-Simons theory with compact gauge group $G$, the Verlinde formula \cite{Verlinde:1988sn} is an explicit expression for $Z_{\mathrm{CS}}[\Sigma\times S^1;G]$ and one of our primary goals in this paper is to obtain its analog --- which we call the ``equivariant Verlinde formula'' --- for Chern-Simons theory with complex gauge group $G_\C$. In contrast to the Verlinde formula that depends on the choice of the gauge group $G$, level $k$, and topology of $\Sigma$, the equivariant Verlinde formula in addition depends also on $\beta$.
Already at this stage one can anticipate some of its properties and behavior in different limits of $\beta$:
\begin{itemize}
\item When $\beta\rightarrow +\infty$, we expect the equivariant Verlinde formula to reduce to the usual Verlinde formula, because in this limit the only contribution to \eqref{CSPartition} comes from the singlet sector with respect to $U(1)_\beta$ and the contributions involving field $\phi$, which is charged under this symmetry, are typically suppressed. Hence, in this limit the $\beta$-deformed complex Chern-Simons theory with gauge group $G_\C$ becomes Chern-Simons theory with compact gauge group $G$. So the equivariant Verlinde formula is a one-parameter deformation of the usual Verlinde formula.
\item When $\beta\rightarrow 0$, we expect the equivariant Verlinde formula to be divergent because in this limit $\beta$ will not provide any regularization for Chern-Simons theory with a complex gauge group $G_\C$.
\end{itemize}
Combining these two points together, one can view the equivariant Verlinde formula as an interpolation between the Verlinde formula with group $G$ and with group $G_\C$.

\subsection{Two different approaches to complex Chern-Simons theory}
\label{sec:twoapproaches}

In general, there are two standard ways to preserve supersymmetry on a curved space $M$:
\begin{itemize}
	\item {\bf Deformation}. One way to preserve supersymmetry is to modify the supersymmetry algebra. An effective way of doing this is to couple the theory to supergravity and find consistent background values for these auxiliary fields \cite{Festuccia:2011ws}. This approach usually 	requires $M$ to have non-trivial isometries.
	\item {\bf Topological Twisting}. Another way is to perform a topological twist \cite{Witten:1988ze}. In a theory realized on world-volume of branes, this operation corresponds to embedding $M$ as a calibrated submanifold in a special holonomy space \cite{Bershadsky:1995qy}. This approach does not require $M$ to have a symmetry.
\end{itemize}

\noindent
Recall our eleven-dimensional geometry \eqref{3d3d2}:
\beq
\begin{matrix}
{\mbox{\rm $N$ fivebranes:}} & \qquad & L(k,1)_b & \times & \Sigma & \times &S^1 \\
& \qquad &   &  & \cap \\
{\mbox{\rm space-time:}} & \qquad & L(k,1)_b & \times & T^* \Sigma & \times & S^1 & \times & \R^3 \\
& \qquad &   &  &\!\!\!\!\!\!\! \circlearrowright &  &  &  &  \circlearrowright \\
{\mbox{\rm symmetries:}} & \qquad &   &   & \!\!\!\! U(1)_N &   &   &   & SU(2)_R
\end{matrix}
\label{3d3d3}
\eeq
We too have two possible ways to formulate the $\beta$-deformed complex Chern-Simons theory living on $\Sigma\times S^1$ as a topological theory with BRST symmetry. The first is to do ``deformation'', which means to reduce 6d $(2,0)$ theory on $L(k,1)$ as in \cite{Cordova:2013cea}, but now in the presence of the $\Omega$-background.
The second (and much easier) approach is to do a topological twist along $L(k,1)$, just like we did it along $M_3$.

In the eleven-dimensional geometry, this can be conveniently achieved by combining the $\R^3$ factor with $L(k,1)$ to obtain $T^*L(k,1)$. As the cotangent bundle of a Lens space is trivial, there is no topological obstruction to doing so, although we do need to modify the metric of $\R^3$ so that the total space is Ricci-flat. In other words, now $L(k,1)$ is embedded as a special Lagrangian submanifold inside a local Calabi-Yau 3-fold:
\beq
\begin{matrix}
{\mbox{\rm $N$ fivebranes:}} & \qquad & L(k,1)_b & \times & \Sigma & \times &S^1 \\
& \qquad & \cap  &  & \cap \\
{\mbox{\rm space-time:}} & \qquad & T^*L(k,1)_b & \times & T^* \Sigma & \times & S^1 \\
& \qquad &  \!\!\!\!\!\!\!\!\!\!\!\!\!\!\!\!\!\!\!\!\!\!\!\circlearrowright &  &\!\!\!\!\!\!\! \circlearrowright   \\
{\mbox{\rm symmetries:}} & \qquad & \!\!\!\!\!\!\!\!\!\!\!\!\!\! SU(2)_R  &   & \!\!\!\! U(1)_N
\end{matrix}
\label{3d3d4}
\eeq
In order to introduce the equivariant parameter $\beta$, we need to single out an $\R^2_{-\beta}$ subspace of $\R^3$ to turn on the $\Omega$-background. So, now we also need to specify how this $\R^2$ is fibered over $L(k,1)$. Lens spaces are particular examples of Seifert manifolds, and $L(k,1)$ is the total space of a degree $k$ $S^1$-bundle over $\bbCP^1$. If we take $\R^2_{-\beta}$ to be the cotangent fiber of $\bbCP^1$, then the two sides of the 3d-3d correspondence are treated on equal footing\footnote{In this paper we focus on the special case $M_3=\Sigma\times S^1$, but it can be replaced with a more general Seifert manifold. And $L(k,1)$ can also be replaced with a more general Seifert manifold, making the two sides of the correspondence completely symmetric.} and this is exactly what we will do.

To summarize, the $\beta$-deformed complex Chern-Simons theory on $\Sigma\times S^1$ can be described as the topological twist of the 3d $\CN=2$ ``Lens space theory'' $T[L(k,1);\beta]$, and our next task is to identify this theory and analyze its dynamics.
Among other things, this gives another possible way to define the graded dimension \eqref{grdimH} or the partition function
of the $\beta$-deformed complex Chern-Simons theory \eqref{CSPartition}:
\beq
\boxed{\phantom{\oint}
\mathrm{dim}_\beta\,\CH_{\mathrm{CS}}(\Sigma;G_\C)
\;=\; Z_{\mathrm{CS}}[\Sigma\times S^1;G_\C, \beta]
\;=\; Z^{\mathrm{twisted}}_{T[L(k,1);\beta]}\left[ \Sigma\times S^1 \right] \,.\phantom{\oint}}
\label{grdimviatwist}
\eeq
In section~\ref{3dTQFT} we present further evidence for the proposed relation~\eqref{3dTQFTsummary} by calculating partition function and
comparing with the prediction of the 3d-3d correspondence, {\it i.e.}~with the equivariant integral over the Hitchin moduli space.


\section{Branes and vortices}
\label{Brane}

The theories studied in this paper describe low-energy physics of certain brane configurations in type IIA and type IIB string theory.
In particular, the type IIB brane configuration will help us identify the Lens space theory $T[L(k,1);\beta]$
and its type IIA dual will make contact with the dynamics of vortices in 4d $\CN=2$ SQCD with a $U(k)$ gauge group.

\subsection{``Lens space theory'' $T\lsb L(k,1)\rsb$ from brane constructions}

\begin{figure}[htb]
	\centering
		\includegraphics[width=0.6\textwidth]{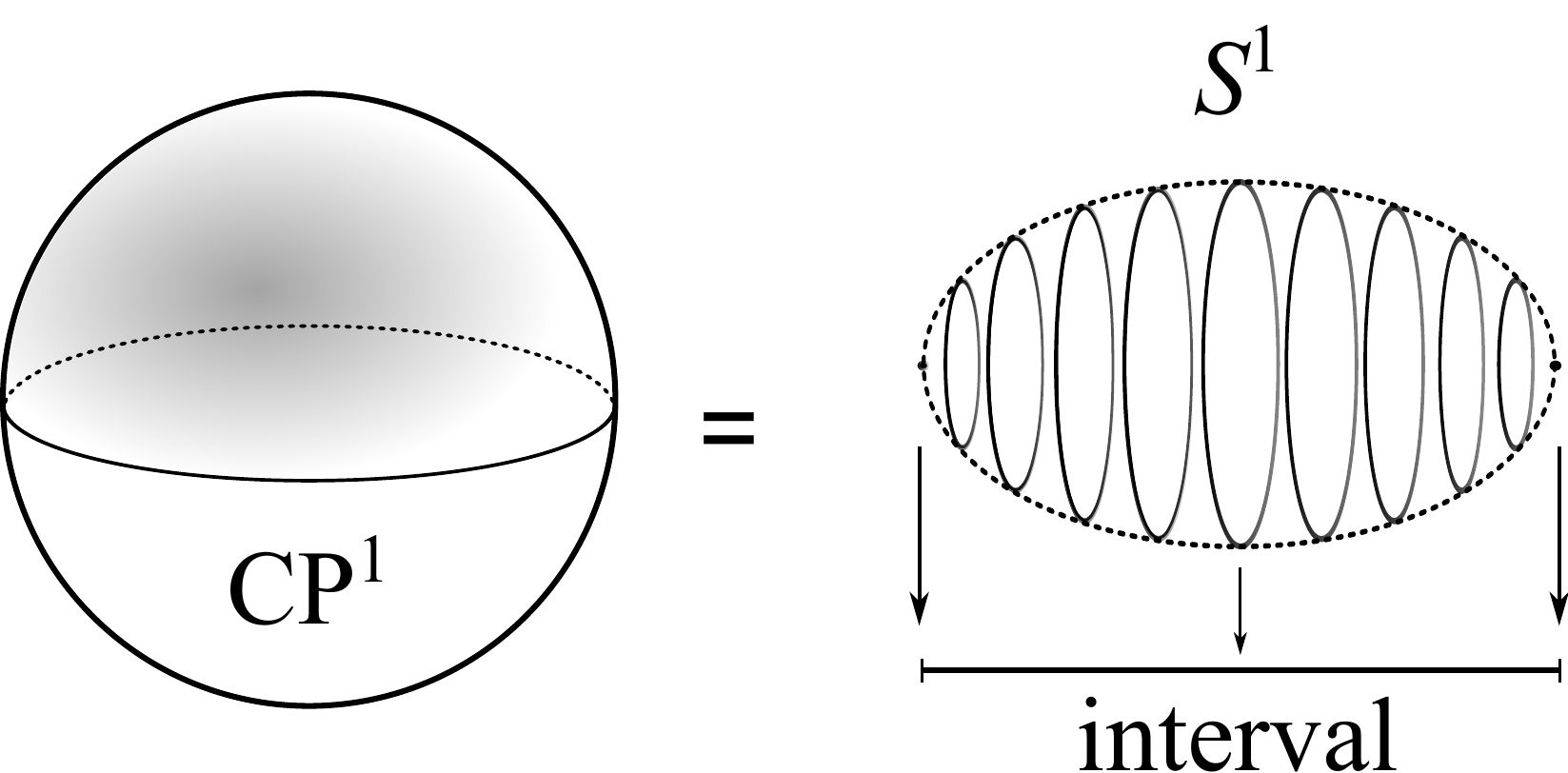}
	\caption{\label{fig:CP1}$\bbCP^1$ can be viewed as the total space of a circle fibration over an interval, with degenerate fibers at the endpoints of the interval.}
\end{figure}

The reduction of the 6d $(2,0)$ theory on $L(k,1)$ can be most easily performed by regarding this Lens space as the total space of a $\T^2$ torus fibration over an interval. At each endpoint of the interval, the torus degenerates to a circle. The first homology group of the torus is
\beq
H_1(\T^2)=\Z\oplus\Z
\eeq
generated by $[a]$ and $[b]$.
Regarding the Lens space $L(k,1)$ as a Hopf fibration over $\bbCP^1$ {\it a la} \eqref{Seifert},
we can also identify $[a]$ with the homology class of the Hopf fiber and $[b]$ with the latitude circle of the base $\bbCP^1$
which shrinks on both ends of the interval, see figure~\ref{fig:CP1}.

\begin{figure}[htb]
	\centering
		\includegraphics[scale=0.85]{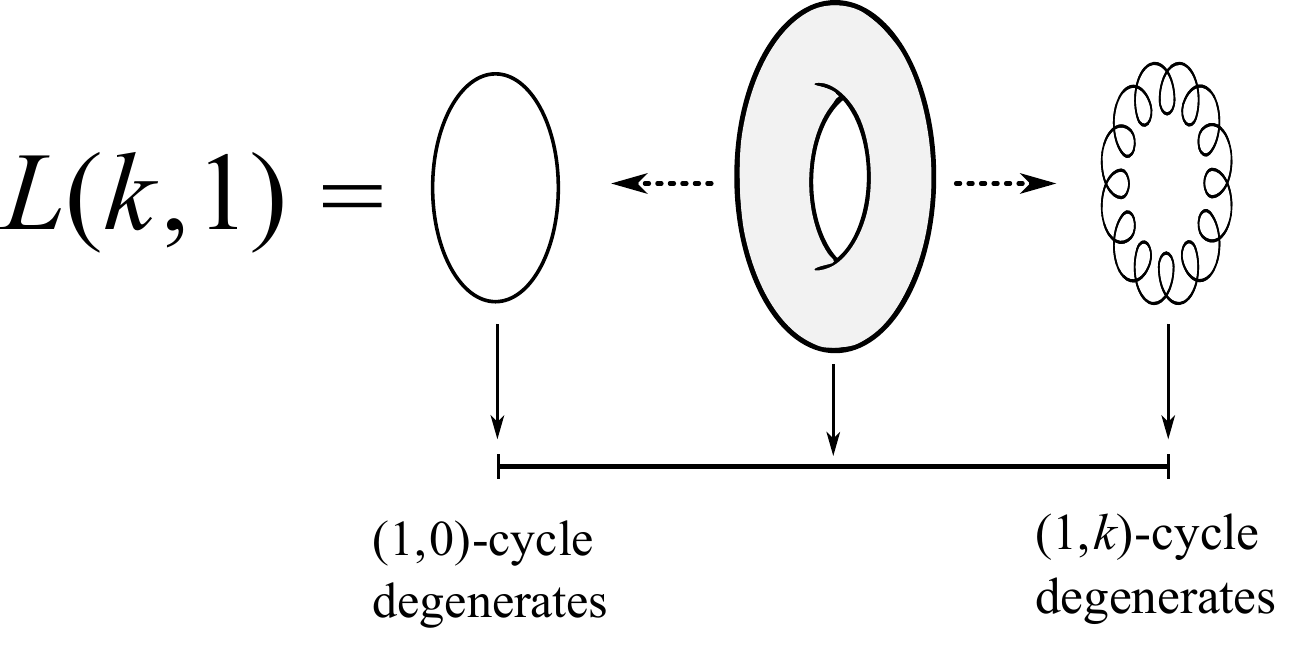}
	\caption{The Lens space $L(k,1)$ can be viewed as the total space of a 2-torus fibered over an interval. Near each endpoint of the interval, a particular cycle of the torus degenerates.}
	\label{fig:LensSpace}
\end{figure}

Then, in representing $L(k,1)$ as a $\T^2$-fibration over the interval, the vanishing cycle at one endpoint of the interval is homologous to $[b]$, whereas non-trivial topology of the Hopf fibration requires the vanishing cycle at the other endpoint of the interval to be $[b]+k[a]$.
This torus fibration is illustrated in figure~\ref{fig:LensSpace}.
Note, near the left endpoint of the interval, the base $\bbCP^1$ looks like a cigar and the total space of its cotangent bundle can be identified with a Taub-NUT space, such that $[b]$ is the $S^1$ fiber that vanishes at the Taub-NUT center.

Now we are ready to reduce our 11-dimensional setup \eqref{3d3d4} on the torus $\T^2$.
Our choice of space-time coordinates is summarized in the following table:
$$
\begin{tabular}{c|c|c|c|c|c|c|c|c|c|c|c}
space-time & 0 & 1 & 2 & 3 & 4 & 5 & 6 & 7 & 8 & 9 & 10 \\
				\hline
     M5 & --- & --- & --- &$\cdot$&$\cdot$&$\cdot$& ---&$\cdot$&$\cdot$&---&---\\
		\hline
geometry& \multicolumn{2}{|c|}{$\Sigma$}& $S^1$ &\multicolumn{2}{|c|}{$\R^2_\beta$}&$\R_{\mathrm{Hopf}}$&Interval& \multicolumn{2}{|c|}{$\R^2_{-\beta}$} & \multicolumn{2}{|c}{$\T^2$}
\end{tabular}
$$
We use $(x^0,x^1,x^2)$ to parametrize $\Sigma\times S^1$, which for now we assume to be flat, until we are ready to implement the topological twist along $\Sigma$. We use $(x^3,x^4)$ to parametrize the cotangent fiber $\R^2_\beta$ of $\Sigma$. And we let the Hopf fiber $S^1$ ($a$-cycle of the torus) to be parametrized by a periodic coordinate $x^9$ and its cotangent space $\R_{\mathrm{Hopf}}$ to be parametrized by $x^5$. We use $x^6$ to be the coordinate on the interval base of the torus bundle, and $(x^7,x^8)$ to be coordinates on the cotangent space $\R^2_{-\beta}$ of $\bbCP^1$, where $\bbCP^1$ is the base of the Hopf fibration. Lastly, we choose the $b$-cycle to be parametrized by $x^{10}$.

\subsubsection*{Type IIB brane configuration}

We are going to use a famous duality between M-theory on a 2-torus and type IIB string theory on a circle,
so that the $SL(2,\Z)$ duality group of type IIB theory has a nice geometric interpretation as the mapping class group of the $\T^2$.
What happens to M5-branes supported on $L(k,1)_b \times \Sigma \times S^1$ upon this reduction?

\begin{figure}[htb]
	\centering
		\includegraphics[scale=0.5]{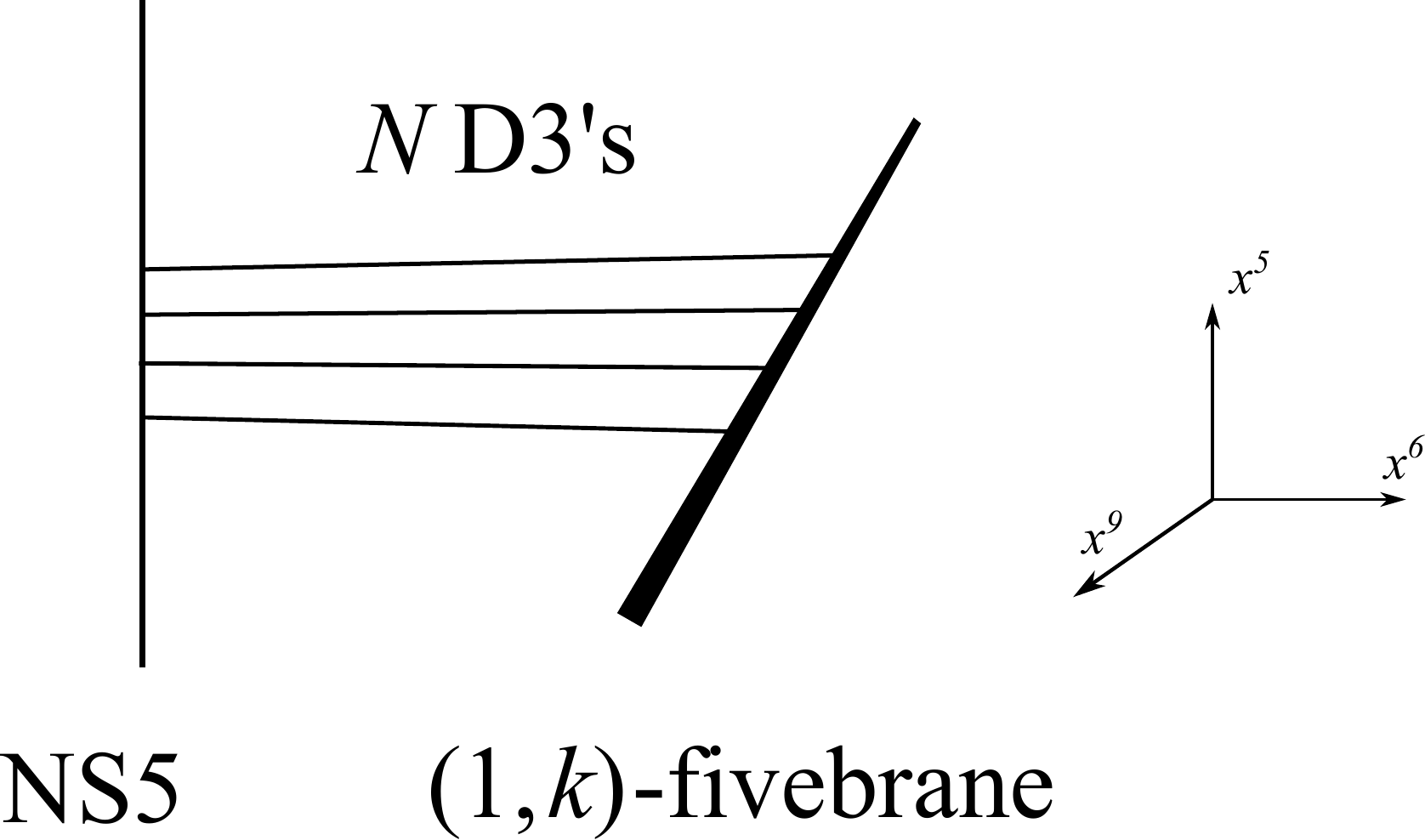}
	\caption{The NS5-D3-(1,$k$) brane system in type IIB string theory.}
	\label{fig:NS5-D3-(1,k)}
\end{figure}

The fivebranes wrapping a torus give rise to a stack of $N$ D3-branes and the boundary condition
it satisfies demands that we have a NS5-brane on one side of the interval and a $(1,k)$-fivebrane
on the other side of the interval:
\beq
\begin{tabular}{c|c|c|c|c|c|c|c|c|c|c}
space-time & 0 & 1 & 2 & 3 & 4 & 5 & 6 & 7 & 8 & 9  \\
				\hline
    $N$ D3's    &---&---&---&$\cdot$&$\cdot$&$\cdot$&$\vdash\!\!\!\dashv$&$\cdot$&$\cdot$&$\cdot$\\
		\hline
     NS5   &---&---&---&---&---&---&$\cdot$&$\cdot$&$\cdot$&$\cdot$\\
		\hline
$(1,k)$-brane&---&---&---&---&---&$\diagdown$&$\cdot$&$\cdot$&$\cdot$&$\diagdown$
\end{tabular}
\label{typeiibbranes}
\eeq
This brane configuration is illustrated in figure~\ref{fig:NS5-D3-(1,k)} and can be equivalently derived as follows.

\begin{figure}[htb]
	\centering
		\includegraphics[scale=0.5]{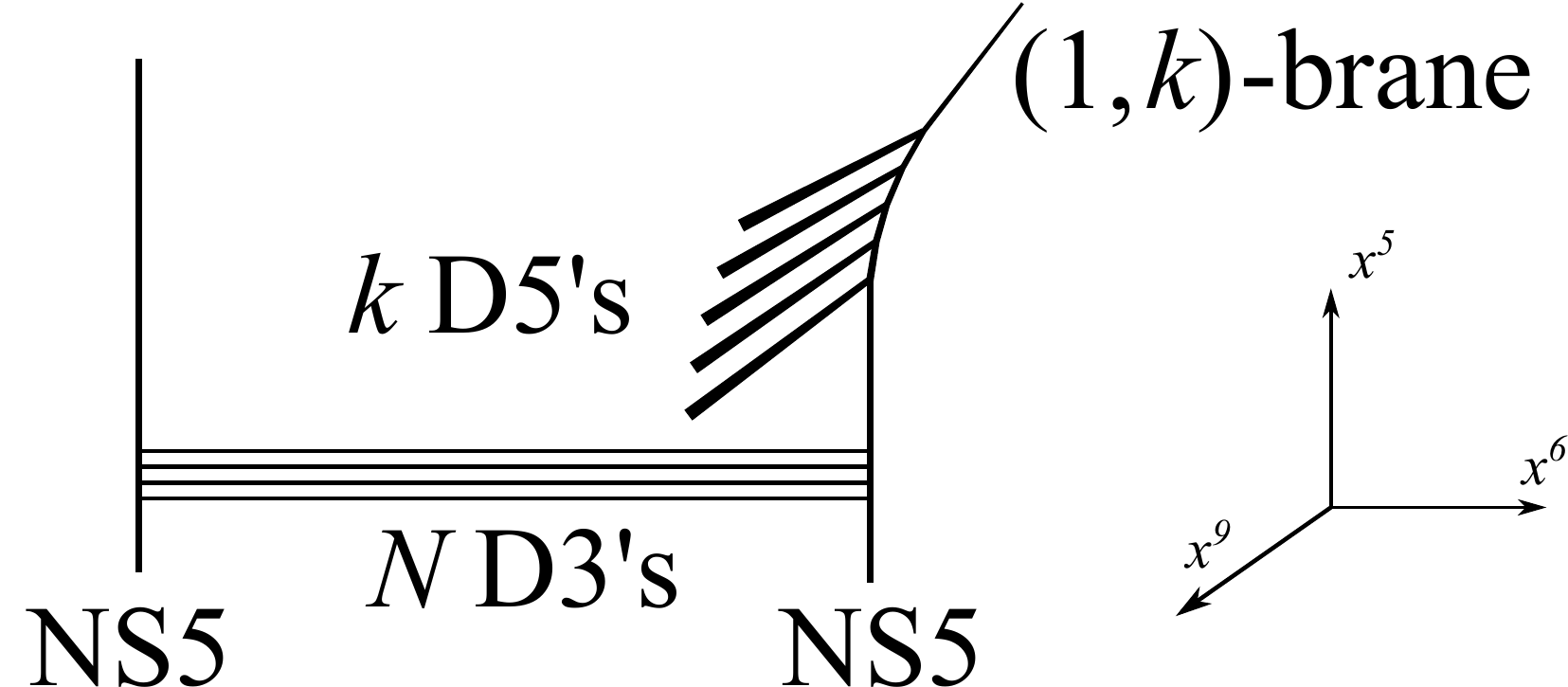}
	\caption{The (1,$k$)-brane in figure~\protect\ref{fig:NS5-D3-(1,k)} is a bound state of an NS5-brane and $k$ D5-branes.}
	\label{fig:NS5-D3-NS5-D5}
\end{figure}

As we pointed out earlier, near the left endpoint of the interval, the base $\bbCP^1$ looks like a cigar and the total space of its cotangent bundle can be identified with a Taub-NUT space, such that $[b]$ is the $S^1$ fiber that vanishes at the Taub-NUT center.
Reducing M-theory on the circle fiber of the Taub-NUT space gives rise to a D6-brane, while $N$ M5-branes become $N$ D4-branes.
In the coordinate system described above, the D6-brane is located at the Taub-NUT center:
\beq
x^6=x^7=x^8=0 \,.
\eeq
In other words, its world-volume spans the space-time directions 0123459. And it is easy to see that the D4-branes are extended along 01269. This is summarized in the table below:
$$
\begin{tabular}{c|c|c|c|c|c|c|c|c|c|c}
space-time & 0 & 1 & 2 & 3 & 4 & 5 & 6 & 7 & 8 & 9  \\
				\hline
     D4    &---&---&---&$\cdot$&$\cdot$&$\cdot$&$\vdash$&$\cdot$&$\cdot$&---\\
		\hline
     D6    &---&---&---&---&---&---&$\cdot$&$\cdot$&$\cdot$&---
\end{tabular}
$$
Here we are looking at the geometry near the left endpoint of the interval $x^6$ so the D4-branes appear to be semi-infinite in the $x^6$ direction.
Then, we perform a T-duality along the Hopf fiber direction parametrized by $x^9$. The D6-brane turns into a D5-brane with world-volume 012345, while the D4-brane becomes a D3-brane spanning 01236. For convenience we perform S-duality, which replaces D5 with an NS5-brane while leaving D3's invariant. We can perform a similar analysis near the right end-point of the interval and obtain $N$ D3's ending on another NS5. But this picture at the right endpoint of the interval is in a different $SL(2,\Z)$ duality frame of type IIB theory; in the original frame we will have a $(1,k)$-fivebrane instead of an NS5-brane. Also, the $(1,k)$-brane is rotated in the $(x^5,x^9)$ plane
\beq
(1,k)\text{-brane}: \quad x^6=x^7=x^8=0,\quad k x^5=x^9,
\eeq
since it can be decomposed into an NS5-brane in 012345 and $k$ D5-branes in 012349, as illustrated in figure~\ref{fig:NS5-D3-NS5-D5}.

To summarize, our M-theory setup \eqref{3d3d4} is dual to the type IIB brane configuration \eqref{typeiibbranes} illustrated in figures~\ref{fig:NS5-D3-(1,k)} and \ref{fig:NS5-D3-NS5-D5}.
In particular, 3d $\CN=2$ Lens space theory $T[L(k,1)]$ can be identified with the theory on D3-branes in figures~\ref{fig:NS5-D3-(1,k)} and \ref{fig:NS5-D3-NS5-D5}.
Besides an 3d $\CN=2$ vector multiplet, it also contains an $\CN=2$ chiral multiplet $\Phi$ in the adjoint representation of the gauge group $G=U(N)$ that corresponds to the motion of D3-branes in directions $x^3$ and $x^4$.
Weakly gauging the $U(1)_\beta$ symmetry \eqref{Uonethetab} that rotates $x^3$ and $x^4$ gives a real mass $\beta$ to $\Phi$:
\beq
\delta S_{\mathrm{mass}} \; = \; \int d^3x d^4\theta \; \Phi \, e^{\beta\theta^2} \,\Phi^\dagger \,.
\label{rephimass}
\eeq
Thus, we end up with the theory described in \eqref{thetwothys}. (Here, $\beta$ plays the role of mass parameter and, hence, is dimensionful. Starting from section~\ref{SUSYSide}, a dimensionless ``equivariant parameter'' $\beta$ will also appear. As they are related simply by a $2\pi R_{S^1}$ factor, with $R_{S^1}$ being the radius of the Seifert $S^1$ fiber, to avoid clutter we use the same symbol $\beta$ for both quantities.)

\subsubsection*{Type IIA brane configuration}

\begin{figure}[htb]
	\centering
		\includegraphics[scale=0.5]{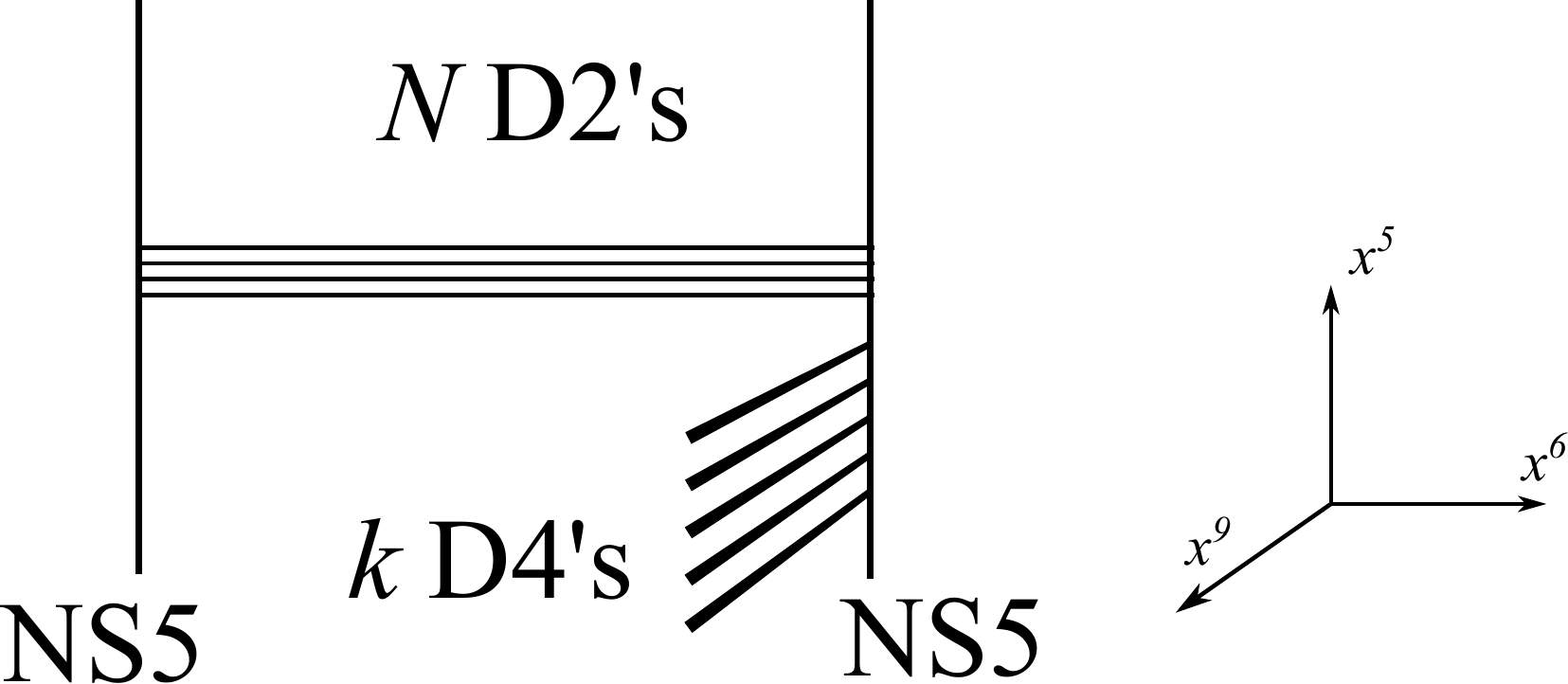}
	\caption{The NS5-D2-NS5-D4 brane system in Type IIA string theory obtained by dimensionally reducing the system in figure~\protect\ref{fig:NS5-D3-(1,k)}.}
	\label{fig:NS5-D2-NS5-D4}
\end{figure}

Our main application of the Lens space theory $T[L(k,1);\beta]$ in this paper is that its twisted partition function on $M_3 = \Sigma \times S^1$ gives the equivariant Verlinde formula. In particular, in sections~\ref{TQFT} and \ref{EVA} we will study the circle reduction of this theory to 2d TQFT on $\Sigma$. The latter is what we are going to call the equivariant $G/G$ gauged WZW model and has a nice interpretation in our brane construction. This dimensional reduction corresponds to a T-duality along the $S^1$ direction parametrized by $x^2$. Upon this T-duality, $N$ D3-branes in figure~\ref{fig:NS5-D3-NS5-D5} transform into $N$ D2-branes in directions 016,
while $k$ D5-branes turn into $k$ D4-branes in directions 01349. The resulting type IIA brane configuration is shown in figure~\ref{fig:NS5-D2-NS5-D4} and describes vortices in $U(k)$ four-dimensional SUSY gauge theory:
\beq
\begin{array}{c}
\text{2d}~\CN=(2,2)~\text{``vortex theory'' on D2-branes} \\[.1cm]
\hline\hline
\text{$U(N)$ SQCD with $k$ fundamental chiral multiplets} \\
\text{and one adjoint chiral multiplet}~\Phi~\text{of mass}~\beta
\end{array}
\label{SQCD}
\eeq

The type IIB and type IIA brane configurations in figures~\ref{fig:NS5-D3-NS5-D5} and \ref{fig:NS5-D2-NS5-D4} will be extremely useful to us for analyzing 3d $\CN=2$ theory $T[L(k,1);\beta]$ and its reduction to 2d, respectively. In particular, we can use either the UV or IR limit of these theories to study topological twist on a Riemann surface $\Sigma$. In the analogous problem that involves 4d $\CN=2$ gauge theory, the twist of the UV theory leads to Donaldson invariants, whereas topological twist of the IR limit leads to Seiberg-Witten equations. Similarly, we can obtain different expressions for the equivariant Verlinde formula (and equivariant Verlinde algebra) by implementing topological twist at different energy scales.

If we perform topological twist in the UV theory, we obtain a 3d TQFT discussed in section~\ref{3dTQFT}. On the other hand, if we follow 3d $\CN=2$ theory $T[L(k,1);\beta]$ to the IR, then we do not even need to perform the topological twist: for generic values of $\beta \ne 0$ the theory has a mass gap and in the IR automatically flows to a TQFT that we call the equivariant $G/G$ gauged WZW model. As we show next, there is yet another phase of the Lens space theory $T[L(k,1);\beta]$ that relates it to a classical problem about vortices.

\subsection{Vortices and equivariant $G/G$ gauged WZW model}
\label{sec:vortices}

Although exactly soluble field theories discussed in this paper have a natural home in mathematical physics, they can be also realized in nature.

In particular, we claim that the low-energy effective theory of $N$ vortices in 4d $\CN=2$ SQCD with $U(k)$ gauge group and $\Omega$-background in the plane orthogonal to the vortex world-sheet is the equivariant $G/G$ model. Furthermore, we claim that the equivariant Verlinde algebra ({\it i.e.} the algebra of loop operators in the $\beta$-deformed complex Chern-Simons theory) whose explicit form will be discussed in section~\ref{EVA} is given by the equivariant quantum K-theory of $\CV_{N,k}$, the moduli space\footnote{Notice, that in the usual notation for the vortex moduli space $k$ stands for the number of vortices, while $N$ is the rank of the gauge group, whose role is reversed in our notations here.} of $N$ $U(k)$ vortices on the plane $\R^2 \cong \C$. Here, the word ``equivariant'' means equivariant with respect to the rotation symmetry of the plane; this is precisely our symmetry $U(1)_{\beta}$. In the physics literature, this equivariant K-theory of vortex moduli spaces was first discussed in \cite{DGH}.

This provides an equivariant generalization of a beautiful story discovered by Witten \cite{Witten:1993xi} that relates the $\Sigma\times S^1$ partition function of $U(N)$ Chern-Simons theory at level $k$ ({\it i.e.} the Verlinde formula) and the algebra of Wilson loops ({\it i.e.} the Verlinde algebra) to the quantum cohomology of the Grassmannian $Gr(N,N+k)$. Our equivariant generalization of this relation can be derived by starting with a ``big theory'':
\bea
& & \text{3d $\CN=2$ $U(N)$ super-Chern-Simons theory at level $\frac{k}{2}$} \nonumber \\
& & + \quad \text{$k$ chiral multiplets $Q_{A=1,\ldots,k}$ in the fundamental representation} \label{bigtheory} \\
& & + \quad \text{1 massive chiral multiplet $\Phi$ in the adjoint representation with mass $\beta$.} \nonumber
\eea
Because the gauge group is $U(N)$, we can turn on an FI parameter $\zeta$ and analyze the vacuum structure as a function of $\zeta$.
We will show that, as $\zeta$ varies, this theory interpolates between the Lens space theory in \eqref{thetwothys} and 3d $\CN=2$ sigma-model with the vortex moduli space $\CV_{N,k}$ as the target and a potential that makes $\CV_{N,k}$ effectively compact.

In order to analyze the vacuum structure of this theory, we need to study the scalar potential as a function of scalar fields, which are the following. Let $\sigma_i$, $i=1,2,\ldots,N$ be the eigenvalues of a scalar field $\sigma$ in the $\CN=2$ vector multiplet. The scalar components of $Q_A$ will be denoted $q^i_{\phantom{i}A}$ and assembled into a $N\times k$ matrix $q$. And the adjoint superfield $\Phi$ contains a $N\times N$ matrix of scalar fields $\varphi_i^{\phantom{i}j}$.

A similar 3d $\CN=2$ theory without the adjoint multiplet $\Phi$ was discussed in \cite{Kapustin:2013hpk}. In the regime $\zeta<0$ it has a unique supersymmetric vacuum where $\sigma$ acquires an expectation value
\beq
\sigma=-\frac{\zeta}{k}\cdot\mathrm{Id} \,.
\eeq
This gives a positive mass to all fundamental chiral multiplets $Q_A$. Integrating these chiral multiplets out leaves us with $\CN=2$ $U(N)$ super-Chern-Simons theory at level $k$. On the other hand, if $\zeta>0$ then one has $\sigma=0$ and the D-term equation is now
\beq
\zeta\cdot\mathrm{Id}=q q^{\dagger} \,.
\eeq
For $k\geq N$, the moduli space of solutions to this equation is the Grassmannian $\mathrm{Gr}(N,k)$ and, therefore, the low energy physics is described by the $\CN=2$ Grassmannian sigma model. If one puts low-energy theories for both $\zeta<0$ and $\zeta>0$ on $\Sigma\times S^1$ and performs the topological twist, one arrives at the conclusion that $U(N)_{k-N}$ Chern-Simons theory\footnote{$\CN=2$ $U(N)_k$ super-Chern-Simons theory is equivalent to $U(N)_{k-N}$ bosonic Chern-Simons theory because integrating out gauginos in the adjoint representation shifts the level by $-N$. } on $\Sigma\times S^1$ is equivalent to the topological A-model of $\mathrm{Gr}(N,k)$. Put in other words, the Verlinde algebra can be identified with the quantum cohomology ring of the Grassmannian. So this argument reproduces the main result of \cite{Witten:1993xi}.

Now let us add the massive adjoint chiral multiplet $\Phi$. For $\zeta\ll 0$, the supersymmetric vacuum characterized by
\beq
\sigma=-\frac{\zeta}{k}\cdot \mathrm{Id}
\eeq
still exists and gives a mass to all the $Q_A$'s. However, the mass of $\Phi$ still comes entirely from \eqref{rephimass}.
Indeed, the only other potential contribution to the mass of its scalar component $\varphi$ is the term
\beq
\big|[\sigma,\varphi]\big|^2 \,,
\eeq
but the identity matrix commutes with any value of $\varphi$. A similar argument shows that fermions in the $\Phi$ multiplet also remain massless as $\sigma$ gets a vev. Therefore, for $\zeta\ll 0$, after integrating out all the fundamental multiplets $Q_A$, the low-energy effective theory is described by $\CN=2$ $U(N)_k$ super-Chern-Simons theory with an adjoint chiral superfield $\Phi$ of (real) mass $\beta$, which is precisely our 3d $\CN=2$ theory $T_N[L(k,1);\beta]$. Hence, we showed that $T_N[L(k,1);\beta]$ can be identified with the $\zeta\ll 0$ phase of the ``big theory'' \eqref{bigtheory}.

On the other hand, in the regime $\zeta>0$ the D-flatness condition of the theory \eqref{bigtheory} looks like
\beq
\zeta\cdot\mathrm{Id}=q q^{\dagger} +[\varphi,\varphi^\dagger].
\eeq
Therefore, the low-energy physics is described by an $\CN=2$ sigma-model with the target space
\beq
\CV_{N,k} \; \cong \; \left\{(q,\varphi)\big|\zeta\cdot\mathrm{Id}=q q^{\dagger} +[\varphi,\varphi^\dagger]\right\}/U(N).
\eeq
This space is conjectured by Hanany and Tong \cite{Hanany:2003hp} to be homeomorphic to the moduli space $\CV_{N,k}$ of $N$ $U(k)$ vortices on $\R^2$. Hence, for $\zeta>0$ the low-energy physics of \eqref{bigtheory} is described by the $\CN=2$ sigma-model with the target space $\CV_{N,k}$ and a potential
\beq
V=\frac{1}{2}\beta^2\left|\varphi\right|^2
\label{vortexV}
\eeq
that comes from the mass of $\Phi$, {\it cf.} \eqref{rephimass}.
Putting the low-energy theories for both $\zeta<0$ and $\zeta>0$ on $\Sigma\times S^1$ and performing the topological twist leads to the following conclusion:
\begin{center}
{\it The $\beta$-deformed complex Chern-Simons theory on $S^1\times \Sigma$ is equivalent to a topological sigma-model to the vortex moduli space $\CV_{N,k}$ equipped with the potential \eqref{vortexV}.}
\end{center}

Note, one can perform different topological twists on $\Sigma$ parametrized by different assignments of the R-charge to the adjoint multiplet $\Phi$. This leads to a large family of quasi-topological theories in three dimensions, only one of which (for $R=2$) happens to be related to complex Chern-Simons theory. It is interesting, though, to study the entire family of such theories, related to different variants of the equivariant quantum K-theory as shown here. Reduction of this family to 2d TQFTs labeled by $R \in \Z$ will be discussed in detail in section~\ref{TQFT}.

It would be interesting to derive the equivariant $G/G$ model on the vortex world-sheet via the anomaly inflow \cite{Callan:1984sa} from 4d $\CN=2$ SQCD with $U(k)$ gauge group (and $\Omega$-background in the plane orthogonal to the vortex world-sheet). A similar question for half-BPS surface operators in 4d gauge theory with $\CN=4$ supersymmetry was studied in \cite{Buchbinder:2007ar}.


\section{Equivariant integration over Hitchin moduli space}
\label{SUSYSide}

In this section we consider the ``supersymmetric'' ({\it i.e.}~left) side of the 3d-3d correspondence \eqref{3d3d0}
when $M_3=\Sigma \times S^1$ or, more generally, a Seifert manifold.
This, in particular, will give the precise meaning to the graded dimension in \eqref{CSPartition}
and show that it can be written as the equivariant integral over the Hitchin moduli space.

As explained in section \ref{fivebranes} and summarized in \eqref{thetwothys}, the 3d $\CN=2$ theory $T[\Sigma \times S^1; \beta]$ is a sigma-model with the Hitchin moduli space $\CM_H$ as the target and has a real mass for the $U(1)_{\beta}$ flavor symmetry, whose action is described in \eqref{Hitbsymm} and \eqref{Uonethetab}.

\subsection{Quantization of Hitchin moduli space}

The dimension of the Hilbert space of Chern-Simons theory with compact gauge group $G$ can be naturally expressed
as an integral over the moduli space of flat connections $\CM_{\mathrm{flat}}$.
Let $A$ be a connection on the principal $G$-bundle over the Riemann surface $\Sigma$ and $F_A$ its curvature. Then the moduli space of flat connections is
\beq
\CM_{\mathrm{flat}}(\Sigma;G)=\left\{A | F_A=0\right\}/\CG,
\eeq
where $\CG$ is the group of gauge transformations. This space is equipped with a natural symplectic form \cite{Atiyah:1982fa}:
\beq
\omega=\frac{1}{4\pi^2}\int_\Sigma \Tr \, \delta A\^\delta A,
\eeq
where $\delta$ is the de Rham differential on $\CM_{\mathrm{flat}}$. With this particular normalization $\omega$ is the generator of the integral cohomology group $H^2(\CM_{\mathrm{flat}},\Z)$.

The classical phase space of Chern-Simons theory at level $k$ on $\Sigma$ is precisely the symplectic space
\beq
\big( \CM_{\mathrm{flat}}(\Sigma;G), k\omega \big),
\label{quantMax}
\eeq
and the Hilbert space $\CH_{\mathrm{CS}}(\Sigma;G,k)$ can be obtained by quantizing it \cite{Witten:1988hf}.
In fact, $\CM_{\mathrm{flat}}(\Sigma;G)$ is a compact K\"aher space as the complex structure of $\Sigma$ defines a complex structure on $\CM_{\mathrm{flat}}(\Sigma;G)$ that is compatible with $\omega$. As a consequence, one can apply the technique of geometric quantization \cite{souriau1966} to identify $\CH_{\mathrm{CS}}(\Sigma;G)$ with the space of holomorphic sections of a ``prequantum line bundle'' $\CL^{\otimes k}$:
\beq
\CH_{\mathrm{CS}}(\Sigma;G,k)=H^0\left(\CM_{\mathrm{flat}}(\Sigma;G),\CL^{\otimes k}\right),
\eeq
where $\CL$ is the universal determinant line bundle with curvature $\omega$.
The index theorem, combined with the Kodaira vanishing theorem for the higher cohomology groups, relates the dimension of the Hilbert space to the index of a spin$^c$ Dirac operator and then to an integral over $\CM_{\mathrm{flat}}(\Sigma;G)$:
\beq
\label{CompactCS}
\mathrm{dim}\,\CH_{\mathrm{CS}}(\Sigma;G,k)=\chi(\CM_{\mathrm{flat}},\CL^{\otimes k})=\mathrm{Index}(\slash \!\!\! \partial_{\CL^{\otimes k}})=\int_{\CM_{\mathrm{flat}}}\mathrm{Td}(\CM_{\mathrm{flat}})\^e^{k\omega},
\eeq
where $\mathrm{Td}\left(\CM_{\mathrm{flat}}(\Sigma;G)\right)$ is the Todd class of $\CM_{\mathrm{flat}}(\Sigma;G)$.

Now let us consider Chern-Simons theory with complex gauge group $G_\C$. The classical phase space is a symplectic manifold
\beq
\big( \CM_{\mathrm{flat}}(\Sigma;G_\C)\cong\CM_H(\Sigma;G),\; k\omega_I+\sigma\omega_K \big) \,.
\label{quantMbx}
\eeq
Here $\CM_H(\Sigma;G)$, later abbreviated as $\CM_H$, is the moduli space of $G$-Higgs bundles over $\Sigma$ \cite{Hitchin:1986vp}:
\beq
\CM_H(\Sigma;G_\C)=\left\{(A,\phi)\left|\begin{array}{rc}
F_A-\phi\^\phi&=0\\
d_A\phi=d_A^\dagger\phi&=0
\end{array}\right.\right\}/\CG.
\eeq
The adjoint-valued one-form $\phi\in\Omega^1(\Sigma,\frak{g})$ is precisely our field $\phi$ that appeared earlier in Table~\ref{Fields}. The Hitchin moduli space is hyper-K\"ahler: it comes equipped with three complex structures $(I,J,K)$ and three real symplectic forms:
\bea
\omega_I&=&\frac{1}{4\pi^2}\int_\Sigma \Tr \left(\delta A\^\delta A-\delta\phi\^\delta\phi\right),\\
\omega_J&=&\frac{1}{2\pi^2}\int_\Sigma \Tr \left(\delta A\^\*\delta \phi\right),\\
\omega_K&=&\frac{1}{2\pi^2}\int_\Sigma \Tr \left(\delta A\^\delta \phi\right) \,.
\eea
This space can be viewed as a natural complexification of $\CM_\mathrm{flat}(\Sigma;G)$ and it is birationally equivalent to $T^*\CM_\mathrm{flat}$.
The canonical determinant bundle $\CL$ also extends naturally to a line bundle over $\CM_H$ that we continue to call $\CL$. The curvature of $\CL$ is now $\omega_I$. (This extension of $\CL$ from $\CM_\mathrm{flat}(\Sigma;G)$ to $\CM_H$ is one of the key elements in the ``brane quantization'' of the moduli space of flat connections \cite{Gukov:2008ve}.)

Just as in the quantization of \eqref{quantMax}, the quantization of \eqref{quantMbx} leads to a Hilbert space whose dimension can be formally expressed as an integral over $\CM_H$ similar to \eqref{CompactCS}:
\beq
\label{ComplexCS1}
\mathrm{dim}\,\CH_{\mathrm{CS}}(\Sigma;G_\C,k)=\int_{\CM_H}\mathrm{Td}(\CM_H)\^e^{k\omega_I+\sigma\omega_K} \,.
\eeq
However, as the Hitchin moduli space is non-compact, the integral above is divergent,
indicating that the Hilbert space associated with complex Chern-Simons theory is infinite-dimensional.

An interesting feature of the Hitchin moduli space is that it admits a circle action with compact fixed point loci which, anticipating a connection with an earlier discussion, we shall call $U(1)_\beta$. This action was used by Hitchin \cite{Hitchin:1986vp} to study topology of the moduli space of Higgs bundles and in the literature is sometimes referred to as ``the Hitchin action''. The corresponding vector field $V$ on $\CM_H$ is generated by the Hamiltonian:
\beq
\mu=\frac{1}{2\pi}\int_\Sigma \Tr(\phi\^\*\phi),
\eeq
with the symplectic form $\omega_I$:
\beq
\delta \mu=2\pi\iota_V\omega_I.
\eeq
Indeed, one can see that this action, rotating the Higgs fields $\phi$, is exactly \eqref{Uonethetab}, which rotates the cotangent space of $\Sigma$ where the field $\phi$ lives.
Using this $U(1)_\beta$ action, we can regularize the divergent integral in \eqref{ComplexCS1} by converting it to an {\it equivariant integral}.
First we define the equivariant differential associated with the Hamiltonian $U(1)_\beta$ action on $\CM_H$:
\beq
D=\delta+2\pi \beta \iota_V \,.
\eeq
Here $\beta$ is the generator of
\beq
H^*_{S^1}(pt)=H^*(\bbCP^{\infty})=\C[\beta] \,,
\eeq
assigned degree 2 in the equivariant cohomology to make $D$ homogeneous. We have chosen $\beta$ for this equivariant parameter, so that it can be identified with the mass parameter in the previous discussion. Then, the equivariantly closed extension of $\omega_I$ is
\beq
\tilde{\omega}_I=\omega_I- \beta \mu,
\eeq
which satisfies
\beq
D\tilde{\omega}_I=0 \,.
\eeq

Because $\omega_K$ is not invariant under $U(1)_\beta$, we set $\sigma$ to zero in \eqref{ComplexCS1}.
In the original problem of quantizing \eqref{quantMbx} it means that we set the ``imaginary part'' of the complex Chern-Simons theory to zero.
Since all the relevant characteristic classes have equivariant extensions,
it is natural to replace the divergent integral (\ref{ComplexCS1}) with an equivariant integral that computes the equivariant index:
\beq\label{ComplexCS2}
\int_{\CM_H} \mathrm{ch}\left(\CL^{\otimes k}\right) \^ \text{Td} (\CM_H)=\mathrm{Index}(\slash\!\!\!\partial_{\CL^{\otimes k}}) \; \leadsto \; \mathrm{Index}_{S^1}(\slash\!\!\!\partial_{\CL^{\otimes k}};\beta)=\int_{\CM_H} \mathrm{ch} \left(\CL^{\otimes k},\beta\right) \^ {\rm Td} (\CM_H, \beta).
\eeq
In particular, the equivariant Chern character
\beq
\mathrm{ch} (\CL^{\otimes k},\beta)=\exp\left(k\tilde{\omega}_I\right)=\exp\left(k\omega_I-k\beta \mu\right)
\eeq
exponentially suppresses the contribution of parts far away from $\CM_\mathrm{flat}(\Sigma; G)\subset\CM_H$, where $\mu\gg 0$.
Therefore, one may hope that a positive value of $\beta$ provides the desired regularization of the naive expression \eqref{ComplexCS1}.

Using the Atiyah-Bott localization formula \cite{MR721448} one can rewrite the right-hand side of (\ref{ComplexCS2})
as an integral over the critical manifolds, $F_d$, of $\mu$:
\beq\label{EVerLocal}
\mathrm{Index}_{S^1}(\slash\!\!\!\partial_{\CL^{\otimes k}},\beta)=\sum_{F_d}e^{- \beta k\cdot \mu(F_d)}\int_{F_d}\frac{\mathrm{Td}(F_d)\^e^{k\omega_I}}{\prod_i\left(1-e^{-x_i-\beta n_i}\right)},
\eeq
which is manifestly convergent as all critical manifolds are compact.
In the denominator we used the splitting principle to decompose the normal bundle of $F_d$
into line bundles $L_i$ whose equivariant Chern classes are $1+x_i+\beta n_i$.

The equivariant index \eqref{EVerLocal} is going to be our definition for the graded dimension
of the Hilbert space of complex Chern-Simons theory \eqref{CSPartition}:
\bea
\mathrm{dim}_\beta\CH(\Sigma;G_\C,k)
& = & Z_{\mathrm{CS}}[\Sigma\times S^1;G_\C,k,\beta] \label{EVer} \\
& =& \mathrm{index}_{S^1}(\slash \!\!\!\partial_{\CL^{\otimes k}};\beta)=\int_{\CM_H} \mathrm{Td}(\CM_H,\beta)\^\exp(k\tilde{\omega}_I) \,. \nonumber
\eea
Note, every quantity in this formula, except for the first one ({\it viz.}~the partition function of complex Chern-Simons theory with $\beta$ deformation) has precise mathematical definition and at this stage can in principle be computed directly. In section~\ref{EVA} we perform the equivariant integration explicitly in the case of $G=SU(2)$ for some punctured Riemann surfaces and obtain the $SU(2)$ ``equivariant Verlinde algebra'' generalizing the usual Verlinde algebra.

However, this direct approach becomes progressively more complicated as the rank of the gauge group gets larger and larger. Our goal is to evaluate \eqref{EVer} indirectly, using the 3d-3d correspondence \eqref{3d3d0} to compactify the fivebrane theory on $L(k,1)$ first and then use string dualities of section~\ref{Brane} to derive the exact solution of the $\beta$-deformed complex Chern-Simons theory on $M_3 = \Sigma\times S^1$ (and, more generally, on Seifert manifolds).
We hope that many alternative ways for computing the integral \eqref{EVer} presented in this paper can shed light on the singularity structure of the moduli space of Higgs bundles (when the rank and the degree are not coprime).

Before we proceed, let us point out that in \cite{Moore:1997dj} a similar integral over $\CM_H$ which computes the ``equivariant volume'',
\beq\label{EVol}
\mathrm{Vol}_\beta(\CM_H)=\int_{\CM_H}\exp(\tilde{\omega}_I),
\eeq
was studied using the ``topological Yang-Mills-Higgs model''. This model was later analyzed in detail in \cite{Gerasimov:2006zt,Gerasimov:2007ap}. As the equivariant index is the K-theoretic lift of the equivariant volume, we expect the $\beta$-deformed complex Chern-Simons theory to share a lot of similarities with the Yang-Mills-Higgs model. In particular, it should have a BRST symmetry. One way to obtain a theory with BRST symmetry is to start with a supersymmetric theory and perform a topological twist. As we will see in the next section, this is indeed the case: the $\beta$-deformed complex Chern-Simons theory on $\Sigma\times S^1$ is equivalent to a topologically twisted 3d $\CN=2$ supersymmetric gauge theory.


\section{$\beta$-deformed complex Chern-Simons}
\label{3dTQFT}

\subsection{Complex Chern-Simons theory from topological twist}
\label{ComplexTwist}

Since generic 3d $\CN=2$ theories have R-symmetry group $U(1)$ they cannot be twisted on general 3-manifolds with holonomy group $SO(3)$. However, if $M_3=\Sigma\times S^1$ is equipped with a metric such that the $U(1)_S$ Seifert action rotating the $S^1$ factor is an isometry, then the holonomy group is reduced to $U(1)$ and one can perform a ``semi-topological'' twist for a 3d $\CN=2$ theory on $M_3$. After the twist, the resulting theory does not depend on the choice of metric, as long as $U(1)_S$ is still an isometry of that metric. Equivalently, upon the dimensional reduction on a circle fiber it gives truly topological theory in two dimensions. When $M_3$ is not $\Sigma\times S^1$ but still Seifert, equipped with a $U(1)_S$ invariant metric, one cannot do the topological twist to a 3d $\CN=2$ theory but can still put it on $M_3$ by deforming the supersymetry algebra. This is the approach taken by K\"all\'en in \cite{Kallen:2011ny} for $\CN=2$ super-Chern-Simons theory and by Ohta and Yoshida in \cite{Ohta:2012ev} for $\CN=2$ Chern-Simons-matter theories. 

Here, we apply this to a particular 3d $\CN=2$ theory, namely $T[L(k,1);\beta]$ that one finds after the reduction of the 6d $(2,0)$ fivebrane theory on a Lens space. As any other 3d $\CN=2$ theory, $T[L(k,1);\beta]$ can be twisted on $\Sigma\times S^1$ or defined on more general Seifert manifolds using deformed SUSY. Then, according to section~\ref{fivebranes}, this theory on $M_3$ will be precisely the sought-after ``$\beta$-deformed $G_\C$ complex Chern-Simons theory'' at level $k$. At this stage, from the definition in section~\ref{fivebranes}, we know the following three facts about this $\beta$-deformed $G_\C$ complex Chern-Simons theory at level $k$:
\begin{enumerate}

\item For $\beta\rightarrow +\infty$ it reduces to Chern-Simons theory with compact gauge group $G$ at level~$k$.

\item For $\beta\rightarrow 0$ it becomes Chern-Simons theory with non-compact gauge group $G_\C$.

\item For general $\beta$, we would expect the theory to produce the equivariant integral \eqref{EVerLocal} over the Hitchin moduli space $\CM_H$ if we put it on $\Sigma\times S^1$.

\end{enumerate}

\noindent
Now we demonstrate that 3d $\CN=2$ theory $T[L(k,1);\beta]$ twisted on $\Sigma\times S^1$ indeed satisfies all these criteria, thereby verifying \eqref{3dTQFTsummary}.
Then, in subsection~\ref{sec:3dlocalization}, we compute its partition function \eqref{grdimviatwist} using localization.

\subsubsection{The limit $\beta\rightarrow +\infty$ and compact group $G$}

In the $\beta\rightarrow +\infty$ limit, the adjoint chiral multiplet $\Phi$ in $T[L(k,1);\beta]$ can be integrated out and it will produce a shift of the Chern-Simons level $k\rightarrow k'=k+h_{\frak{g}}$, where $h_{\frak{g}}$ is the dual Coxeter number of the Lie algebra $\frak{g}$. Then we are left with $\CN=2$ super-Chern-Simons theory with gauge group $G$ at level $k'$. This theory can be further reduced to pure bosonic Chern-Simons theory after integrating out gauginos $\lambda, \lambda^\dagger$ and bosonic fields $\sigma, D$. The functional determinant associated with gauginos is not well defined and one needs to regularize it. A standard way to do this is to add a Yang-Mills term to the theory and send the Yang-Mills coupling to infinity. Using this regularization, which is natural from the brane picture, the functional integral over gaugino fields will produce a further shift $k'\rightarrow k$, see e.g. \cite{Kao:1995gf}.

Notice that expectation values of physical observables in Chern-Simons theory at level $k$ usually depend on $k'=k+h_{\frak{g}}$, which comes from gluon loops. Combined with this, there are in total three level-shifting effects, which are summarized below.
\begin{enumerate}
	\item Integrating out $\CN=2$ adjoint chiral multiplet with large positive mass shifts the level by $+h_{\frak{g}}$.
	\item Integrating out gauginos in super-Chern-Simons theory shifts the level by $-h_{\frak{g}}$.
	\item Integrating over gauge fields to compute partition function or expectation values of physical observables effectively renormalizes the level by $+h_{\frak{g}}$.
\end{enumerate}

The effects of 1 and 2 cancel each other so that $T[L(k,1);\beta\rightarrow \infty]$ is equivalent to pure bosonic Chern-Simons theory at level $k$.

\subsubsection{The limit $\beta \rightarrow 0$ and complex group $G_{\C}$}

In this limit $T[L(k,1);\beta]$ is a superconformal theory and topological twist is crucial in order to produce a TQFT. (In general, a gapped theory is expected to flow to a TQFT in the infrared even without a topological twist.) The topological twist of $\CN=2$ super-Chern-Simons theory with general matter content on a Seifert manifold is discussed in \cite{Ohta:2012ev}. In particular, on $\Sigma\times S^1$ a chiral multiplet will yield two BRST-multiplets $(\varphi,\psi)$ and $(\chi,\eta)$. Here $\varphi$ and $\eta$ are bosons while $\psi$ and $\chi$ are fermions. Regarded as fields on $\Sigma$, they are respectively sections of
\bea
(\varphi,\psi)&\in&\Gamma\left[\Omega^0(L\frak{g}\otimes \C)\right],\\
(\chi,\eta)&\in&\Gamma\left[\Omega^1(L\frak{g})\right], \nonumber
\eea
where $L\frak{g}$ is the Lie algebra of the loop group $LG$. Using the complex structure of the Riemann surface, one can decompose $(\chi,\eta)$ into $(1,0)$-forms $(\chi_z,\eta_z)$ and $(0,1)$-forms $(\chi_{\bar{z}},\eta_{\bar{z}})$. Similarly, the components of a vector multiplet $(A, \lambda, \sigma, D)$ now become $(A_z, A_{\bar{z}}, A_0, \lambda_{z}, \lambda_{\bar{z}}, \lambda_0, \sigma, D)$. (See appendix of \cite{Ohta:2012ev} for definitions of these fields and their transformation rules.) In what follows, we will focus on the matter part which comes from the chiral multiplet $\Phi$. The corresponding BRST transformations are\footnote{Notation in \cite{Ohta:2012ev} differs from ours by $z\leftrightarrow \bar{z}$. The notation used here is chosen to agree with that in gauged WZW-matter model, which will be discussed below.}
\bea
Q\varphi=\psi, & Q\psi=-i\CD_0\varphi-i\sigma\varphi \,,  \nonumber\\
Q\chi_z=\eta_z, & Q\eta_z=-i(\CD_0+\sigma)\chi_z +\beta\chi_z \,, \\
Q\chi_{\bar{z}}=\eta_{\bar{z}}, & Q\eta_{\bar{z}}=i(\CD_0+\sigma)\chi_{\bar{z}} +\beta\chi_{\bar{z}} \,.  \nonumber
\eea

However, this is not the only possible twist of the 3d $\CN=2$ theory $T[L(k,1);\beta]$. The twist described above corresponds to assigning R-charge\footnote{Our convention is such that the superspace coordinates $\theta$ has R-charge 1.} $R=0$ for $\Phi$. Since the new Lorentz group of the Riemann surface $U(1)'_L$ is taken to be the diagonal subgroup of $U(1)_L\times U(1)_R'$, this assignment makes the scalars $\varphi$ remain scalar after the twist. As $T[L(k,1);\beta]$ has no F-term interactions\footnote{Recall, that the real mass is given by a D-term.} and the $U(1)_R'$ R-charge assignment for $\Phi$ is unconstrained, nothing prevents us from considering more general integer values of $R$. In particular, what turns out to be related to complex Chern-Simons theory is the case of $R=2$. When $R=2$, the fields are sections of:
\bea
(\varphi,\psi)&\in&\Gamma\left[\Omega^1(L\frak{g})\right] \,, \\
(\chi,\eta)&\in&\Gamma\left[\Omega^0(L\frak{g}\otimes \C)\right] \,, \nonumber
\eea
and we will write them in components as $(\varphi_z,\varphi_{\bar{z}},\psi_z,\psi_{\bar{z}},\chi,\eta)$. The BRST transformations are: 
\bea
Q\varphi_z=\psi_z, & Q\psi_z=-(\CD_0+\sigma)\varphi_z +\beta\varphi_z \,,  \nonumber \\
Q\varphi_{\bar{z}}=\psi_{\bar{z}}, & Q\psi_{\bar{z}}=(\CD_0+\sigma)\varphi_{\bar{z}} +\beta\varphi_{\bar{z}} \,, \\
Q\chi=\eta, & Q\eta=-i(\CD_0+\sigma)\chi \,.  \nonumber
\eea

Now we describe the relation between this twisted SUSY theory and complex Chern-Simons theory, whose action at level $(k,\sigma) = (k,0)$ is
\beq
\label{CSAction}
S_{\mathrm{CS}}^{(k,0)}(\underline{A},\underline{\phi})=\frac{k}{4\pi}\int\left(\underline{A}\^d\underline{A}+\frac{2}{3}\underline{A}\^\underline{A}\^\underline{A}-\underline{\phi}d_{\underline{A}}\underline{\phi}\right),
\eeq
where $\underline{A}=A + A_0 dx^0$ and $\underline{\phi}=\phi + \phi_0 dx^0$ are gauge fields in 3d. We see that the part involving Higgs field $\phi$, which will eventually be identified with the adjoint scalar Lagrangian in $T[L(k,1);\beta=0]$, is well separated from the gauge field Lagrangian.

At this stage, there are two obvious disconnects with the twist of $\CN=2$ theory $T[L(k,1)]$. First of all, the $U(1)_\beta$ flavor symmetry is missing in complex Chern-Simons theory. Secondly, complex Chern-Simons theory is invariant under a larger gauge group $G_\C$. The two difficulties actually cancel each other as we will see next.

We first rewrite the action (\ref{CSAction}) in the geometry $\Sigma\times S^1$:
\beq
S_{\mathrm{CS}}^{(k,0)}(A,\phi,A_0,\phi_0)=\frac{k}{4\pi}\int_{\Sigma\times S^1}\Tr\left(A\^D_0 A+2A_0\^A\^A+2A_0\^dA-2\phi_0\^d_A\phi-\phi \^D_0\phi\right).
\eeq
Here $D_0$ is the covariant derivative along the $S^1$ fiber of the Seifert manifold or $\Sigma\times S^1$ in our basic example. The integral over $\phi_0$ can be explicitly carried out and gives a delta function that implements the constraint
\beq
d_A \phi=0 \,.
\eeq
After integrating out $\phi_0$, the Lagrangian is invariant under $U(1)_{\beta}$, but the condition above is not. A natural way to cure this problem is to impose the gauge choice
\beq
d_A^\dagger \phi=0 \,.
\eeq
Note, the above two equations are also two of the three Hitchin equations. After these steps, the only term in the Lagrangian that depends on $\phi$ is proportional to
\beq
\phi\^D_0\phi \,.
\eeq

In the twisted 3d $\CN=2$ theory $T[L(k,1);\beta=0]$, the whole matter part of the action is $Q$-exact, and nothing prevents us from changing it into another $Q$-exact term, such as
\beq
\frac{1}{2}Q\left(\varphi_z\^\psi_{\bar{z}}-\psi_z\^\varphi_{\bar{z}}\right)=\psi_z\^\psi_{\bar{z}}+\varphi_{z}\^D_0\varphi_{\bar{z}}.
\eeq
After integrating out $\psi$, gauginos $\lambda$, scalars $\sigma$ and $D$, we obtain precisely the complex Chern-Simons action. (Notice, that the shifts of level caused by $\psi$ and $\lambda$ cancel each other.)

\subsection{Equivariant Verlinde formula}
\label{sec:3dlocalization}

The Verlinde formula is usually written as a sum over highest weight integrable representations of the loop group $LG$ at level $k$
(see {\it e.g.} \eqref{usualVerSU2} for $G=SU(2)$, in which case it is simply a sum over $j = 1,2, \ldots, k+1$):
\beq
j ~\in~ \Lambda_{G,k} = \left( \frac{\Lambda_{\mathrm{wt}}}{W \times (k+h) \Lambda_{\mathrm{rt}}} \right)' \,.
\eeq
Here $h_{\frak{g}}$ is the dual Coxeter number of the Lie algebra $\frak{g}$ and the prime means that the fixed points are removed.
It is natural to expect that the equivariant Verlinde formula, defined as the partition function of the $\beta$-deformed complex Chern-Simons theory \eqref{CSPartition}, takes a similar form.

Now, once we established the equivalence of the $\beta$-deformed complex Chern-Simons with the twist of 3d $\CN=2$ theory $T[L(k,1);\beta]$ described in the previous subsection, we can use the standard localization techniques to compute its partition function.
Thus, one can follow {\it e.g.} the techniques of \cite{Ohta:2012ev} to calculate the partition function
of the $\beta$-deformed complex Chern-Simons theory not only on $\Sigma\times S^1$ but on any Seifert manifolds $M_3$,
and with arbitrary R-charge assignment for adjoint chiral multiplet $\Phi$.
Here, for simplicity, we focus on the particular case of $R=2$ and $M_3=\Sigma\times S^1$.
Generalization of the equivariant Verlinde formula to arbitrary value of $R \in \Z$ will be discussed in the next section from a 2d perspective.

Using the localization procedure described in \cite{Ohta:2012ev}, one can express the whole partition function as a path integral over two-dimensional abelian fields
\begin{multline}\label{3dPathInt}
Z^{\beta-\text{CS}}\left[\Sigma;U(N),k,t\right]=\frac{1}{|W|}\int\CD\sigma_a\CD \lambda_a\CD A_a \left[\prod_\alpha\left(1-e^{2\pi i (\sigma_a-\sigma_b)}\right)^{1-h}\right]\Xi^{\text{3d}}\, \\
\times  \exp\left\{i\int_\Sigma\left[ \left((k+N) \sigma_a-\sum_{b=1}^N\sigma_b+\frac{N-1}{2}\right) F^a+\frac{k}{4\pi}\lambda_a \^ \lambda_a\right]\right\},
\end{multline}
where $(\sigma_a,\lambda_a, A_a), a=1,2,\ldots,N$ are fields living on $\Sigma$ and valued in the Cartan of $\frak{u}(N)$.
The important factor $\Xi^{\text{3d}}$ is the matter contribution to the path integral
\beq\label{3dDet}
\Xi^{\text{3d}}=\frac{\det_\chi\left[-i\CL_0-\frac{\mathrm{ad}(2\pi\sigma)+i\beta}{\ell}\right]}{\det_\varphi\left[-i\CL_0-\frac{\mathrm{ad}(2\pi\sigma)+i\beta}{\ell}\right]},
\eeq
where $\CL_0$ is the Lie derivative along the Seifert fiber and $\ell=2\pi R_{S^1}$ is the circumference of the Seifert $S^1$ fiber. If we set $\Xi^{\text{3d}}$ to a constant by sending $\beta$ to infinity, the rest of the path integral is exactly the partition function of Chern-Simons theory on $\Sigma\times S^1$ and it gives the ordinary Verlinde formula. Hence, the functional determinant \eqref{3dDet} contains interesting information about how the equivariant Verlinde formula depends on the deformation parameter $\beta$ and we now evaluate it.

First we decompose $\chi$ and $\varphi$ into Fourier modes
\bea
\chi(z,\bar{z},\theta)&=&\sum_{m\in\Z}\chi_m(z,\bar{z})e^{-im\theta}, \\
\varphi(z,\bar{z},\theta)&=&\sum_{m\in\Z}\varphi_m(z,\bar{z})e^{-im\theta}. \nonumber
\eea
These modes are sections of
\bea
\chi_m   &\in& \Gamma[\Omega^0(\Sigma, \frak{g}\otimes \C)],  \\
\varphi_m&\in& \Gamma[\Omega^1(\Sigma, \frak{g})].  \nonumber
\eea
Then (\ref{3dDet}) can be decomposed into
\beq
\prod_{m\in\Z}\frac{\det_\chi[-i\CL_0-\mathrm{ad}\left(\frac{2\pi\sigma}{\ell}\right)-\frac{i\beta}{\ell}]}{\det_\varphi[-i\CL_0-\mathrm{ad}\left(\frac{2\pi\sigma}{\ell}\right)-\frac{i\beta}{\ell}]}=\prod_{\alpha}\prod_{m\in\Z}\left[-\frac{2\pi m}{\ell}-\alpha\left(\frac{2\pi\sigma}{\ell}\right)-\frac{i\beta}{\ell}\right]^{\mathrm{Index}\,\bar{\partial}_A|_{(\alpha)}}.
\eeq
Here $\alpha$ runs over all roots of $\frak{g}$. From this expression, it is easy to see that $\ell$ only enters as a normalization factor, in agreement with the TQFT nature of the $\beta$-deformed complex Chern-Simons theory.

After ignoring a normalization factor that does not depend on the deformation parameter $\beta$, the functional determinant is
\beq
\Xi^{\text{3d}}=\prod_{\alpha}\left\{\left(\alpha(2\pi\sigma)+i\beta\right)\prod_{m=1}^{+\infty}\left[(2 \pi m)^2-\left(\alpha(2\pi\sigma)+i\beta\right)^2\right]\right\}^{1-h-\alpha(n)}.
\eeq
Here we also used the index theorem
\beq
\mathrm{Index}\,\bar{\partial}_A|_{(\alpha)}=1-h-\alpha(n),
\eeq
with the last term being the degree of the line bundle labeled by $\alpha$:
\beq\label{Redun1}
\alpha(n)=\frac{1}{2\pi}\int_\Sigma\alpha_a F^a.
\eeq
The infinite product over $m$ gives a sine function:
\begin{multline}
\Xi^{\text{3d}}=\prod_{\alpha}\left[\left(\alpha(2\pi\sigma)+i\beta\right)\prod_{m=1}^{+\infty} (2\pi m)^2 \cdot\left(1-\frac{\left(\alpha(2\pi\sigma)+i\beta\right)^2}{(2\pi m)^2}\right)\right]^{1-h-\alpha(n)}
\\
\propto \prod_{\alpha}\left[2\sin\left(\alpha(\pi\sigma)+\frac{i\beta}{2}\right)\right]^{1-h-\alpha(n)}=\prod_{\alpha}\left|1-e^{2\pi i\alpha(\sigma)-\beta}\right|^{1-h-\alpha(n)}.
\end{multline}
Introducing $t=e^{-\beta}$, we decompose the contribution of abelian fields (product over zero roots in $\prod_{\alpha}$) from that of non-abelian fields (product over non-zero roots):
\beq
\Xi^{\text{3d}}=\Xi^{\text{3d}}_{\text{ab}}\cdot \Xi^{\text{3d}}_{\text{nab}},
\eeq
where the abelian functional determinant, modulo a normalization factor\footnote{We did not keep track of the overall normalization constant, but it can be easily restored by demanding that $\beta\rightarrow +\infty$ gives back the usual Verlinde formula.}, is given by
\beq
\Xi^{\text{3d}}_{\text{ab}}=\frac{1}{(1-t)^{N(h-1)}},
\eeq
while the non-abelian contribution is
\beq
\Xi^{\text{3d}}_{\text{nab}}=\left[\prod_{\alpha\neq 0}M_\alpha(\sigma,t)\right]^{1-h-\alpha(n)},
\eeq
with
\beq
M_\alpha(\sigma,t)=1-te^{2\pi i \alpha(\sigma)}.
\eeq
The non-abelian contribution $\Xi^{\text{3d}}_{\text{nab}}$ can be further decomposed into
\beq
\Xi^{\text{3d}}_{\text{nab}}=\prod_{\alpha\neq 0}\left[M_\alpha(\sigma,t)\right]^{1-h}\cdot e^{-\frac{1}{2\pi}\int_\Sigma \alpha(F)\log M_\alpha}.
\eeq
The part that depends on $F$ can be combined with another term in \eqref{3dPathInt}:
\beq
i\int_\Sigma\left((k+N) \sigma_a-\sum_{b=1}^N\sigma_b+\frac{N-1}{2}\right) F^a
\eeq
to give
\beq
i\int_\Sigma\zeta_a F^a,
\eeq
where
\beq
\zeta_a(\sigma)=k\sigma_a-\frac{i}{2\pi}\sum_{b\neq a}\log \left(\frac{e^{2\pi i \sigma_a}-te^{2\pi i \sigma_b}}{te^{2\pi i \sigma_a}-e^{2\pi i \sigma_b}}\right).
\eeq
Performing a functional integral over $A_a$ and over non-zero modes of $\lambda_a$ in \eqref{3dPathInt} gives a collection of delta-functions requiring $\zeta_a$ to be an integer:
\beq
\sum_{l_a\in \Z}\delta(\zeta_a-l_a).
\eeq
Then we integrate over $\sigma_a$'s. The delta-functions produce a factor
\beq
\sum_{\{\sigma\}\in\{\mathrm{Bethe}\}}\det\left| \frac{\partial \zeta_a}{\partial \sigma_b}\right|^{-1}.
\eeq
Here $\{\mathrm{Bethe}\}$ stands for the set of solutions to the following Bethe ansatz equations:
\beq\label{Bethe}
e^{2\pi i k\sigma_a}\prod_{b\neq a}\left(\frac{e^{2\pi i \sigma_a}-te^{2\pi i \sigma_b}}{te^{2\pi i \sigma_a}-e^{2\pi i \sigma_b}}\right)=1,\quad \text{for all of $a=1,2,\ldots,N$}.
\eeq
The set of solutions to the Bethe ansatz equations is acted upon by the Weyl group, and after modding out by this symmetry, the solutions are labeled by Young tableaux with at most $N$ rows and $k$ columns. Notice that the Bethe ansatz equations are the same for all choices of the R-charge assignment to the adjoint chiral multiplet $\Phi$.

Further integrating over the zero modes of $\lambda_a$ gives a factor
\beq
\left|\frac{\partial \zeta_a}{\partial \sigma_b}\right|^{h}.
\eeq
Therefore, the partition function is
\beq
Z^{\beta-\text{CS}}(\Sigma;U(N),k,t)=\sum_{\{\sigma\}\in\{\mathrm{Bethe}\}}\left[\frac{1}{(1-t)^N }\det\left| \frac{\partial \zeta_a}{\partial \sigma_b}\right|\prod_{a\neq b}\frac{1}{\left(e^{2\pi i \sigma_a}-te^{2\pi i \sigma_b}\right)\left(e^{2\pi i \sigma_a}-e^{2\pi i \sigma_b}\right)}\right]^{h-1}.
\label{Z3dend}
\eeq
This ``equivariant Verlinde formula'' enjoys many interesting properties, some of which extend the remarkable properties of the ordinary Verlinde formula, {\it cf.} \eqref{usualVerSU2}. In the next section, we present yet another derivation of this formula, from the two-dimensional point of view. Furthermore, we extend it to an entire family parametrized by the choice of the R-charge assignment of $\Phi$ and then make various comments about this general result.


\section{A new family of 2d TQFTs}
\label{TQFT}

In the previous section, we have seen that twisted 3d $\CN=2$ theory $T[L(k,1);\beta]$ on $\Sigma\times S^1$ can be viewed as a one parameter deformation of complex Chern-Simons theory and it provides a natural way to regularize the latter theory.
In fact, there is an entire family of twisted theories labeled by $R \in \Z$, the R-charge of the adjoint multiplet $\Phi$ in 3d $\CN=2$ theory $T[L(k,1);\beta]$.

In this section, we wish to study dimensional reduction of this family to two dimensions. In particular, we find a new family of 2d TQFTs labeled by $R \in \Z$ that generalize the $G/G$ gauged WZW model and compute their partition functions on an arbitrary Riemann surface $\Sigma$. In certain special cases, we can compare our results to the previous literature.

\subsection{Equivariant $G/G$ gauged WZW model}
\label{sec:equivariantGG}

We know from section~\ref{sec:vortices} that the low-energy dynamics of $T[L(k,1);\beta]$ is given by a topological sigma-model to the vortex moduli space with a potential. In the limit $\beta\rightarrow +\infty$, the effective target space of the sigma-model becomes the Grassmannian and the topological sigma-model is equivalent to the $G/G$ gauged WZW model. Our next goal is to give an equivariant generalization of the gauged WZW model, which we call the ``equivariant $G/G$ gauged WZW model''.

The Lagrangian formulation of this theory can be directly obtained by dimensional reduction of the $\beta$-deformed complex Chern-Simons theory on $S^1$, but we won't follow this approach. Instead, we write down the Lagrangian formulation of the equivariant $G/G$ gauged WZW model and then show that it leads to the same partition function on $\Sigma$ as the $\beta$-deformed complex Chern-Simons theory on $\Sigma\times S^1$.

The fields in the ordinary, non-equivariant $G/G$ model are $(A, \lambda, g)$, where $A$ is the gauge field, $g\in\CG\cong\mathrm{Map}(\Sigma,G)$ is a group-valued field, and $\lambda$ is an auxiliary Grassmann 1-form in the adjoint representation that is required to make the BRST symmetry manifest. The BRST charge $Q_g$ depends on $g$ and takes the following form \cite{Blau:1994et}:
\bea
Q_g A&=&\lambda, \nonumber \\
Q_g \lambda^{(1,0)}&=&\left(A^g\right)^{(1,0)}-A^{(1,0)},\\
Q_g \lambda^{(0,1)}&=&-\left(A^{g^{-1}}\right)^{(0,1)}+A^{(0,1)}, \nonumber
\eea
where
\beq
A^g=g^{-1}A g+g^{-1}d g.
\eeq
At level $k$, the action of the $G/G$ model is
\beq
kS_{G/G}(A,\lambda, g)=k S_G(A,g)-ik\Gamma(A,g)+\frac{i}{4\pi}\int_\Sigma \Tr(\lambda\^\lambda),
\eeq
with the first term on the right-hand side being the kinetic term
\beq
S_G(g,A)=-\frac{1}{8\pi}\int_\Sigma \Tr(g^{-1}d_A g\^ \* g^{-1}d_A g),
\eeq
and the second term being the topological term
\beq
\Gamma(g,A)=\frac{1}{12\pi}\int_B \Tr\left[\left(g^{-1}dg\right)^3\right]-\frac{1}{4\pi}\int_\Sigma \Tr\left(Adg g^{-1}+A A^g\right).
\eeq
Here, $B$ is a handlebody with $\partial B=\Sigma$.

Now we add the chiral multiplet
\beq
\Phi=\varphi+\theta_{\pm}\psi^{\pm}+\theta^2 F,
\eeq
and perform the topological twist. In order to do this, just like in three dimensions, we need to assign R-charge $R$ to the superfield $\Phi$ under $U(1)_V$. The brane construction discussed in section \ref{Brane} naturally leads to $R=2$, but one can consider more general situations, where $R$ is an arbitrary integer.

Identifying the diagonal subgroup of $U(1)_L\times U(1)_V$ with the twisted Lorentz group makes $\varphi$ a section of $\Omega^0(\Sigma, K^{R/2})$, $\psi^{\pm}$ a section of $\Omega^0\left(\Sigma, K^{(R-1\pm 1)/2}\right)$, and $F$ a section of $H^0(\Sigma,K^{R/2-1})$, where $K$ is the canonical bundle of the Riemann surface $\Sigma$. So, after the twist we end up with two BRST-multiplets that come from $\Phi$:
\bea
(\varphi,\psi=\psi^{+}) &\in& \Gamma\left[\Omega^0(\Sigma,K^{R/2})\right], \label{Spin} \\
(\chi=\psi^{-}, \eta=F) &\in& \Gamma\left[\Omega^0(\Sigma,K^{R/2-1})\right] \,, \nonumber
\eea
along with their complex conjugate $(\varphi^\dagger,\psi^\dagger)$ and $(\chi^\dagger,\eta^\dagger)$ from $\Phi^\dagger$.

For $R=2$, the fields $(\chi,\eta)$ are scalars while $(\varphi,\psi)$ are $(1,0)$-forms, which indeed corresponds to the geometry of M5-branes wrapped on $\Sigma \subset T^*\Sigma$. Similarly, for $R=0$, the fields $(\varphi,\psi)$ are scalars, while $(\chi,\eta)$ are $(0,1)$-forms. This choice of the R-charge corresponds to the geometry of $\Sigma \times \C$. We come back to the detailed discussion of these two choices after describing the family of 2d TQFTs labeled by arbitrary (even) integer values of $R$.

At this stage, one can proceed in many different ways to study this family of TQFTs parametrized by $R$. For example, one can take a ``top-down approach'' by starting with the UV Lagrangian of the $\CN=(2,2)$ SQCD with a massive adjoint chiral superfield $\Phi$ and study the resulting model after topological twist using localization.\footnote{An example of this theory, for $R=2$, is the world-volume theory on D2-branes in figure~\ref{fig:NS5-D2-NS5-D4}.} However, since our goal is to generalize the gauged WZW model, we would like to have an explicit Lagrangian formulation that resembles the gauged WZW model. In fact, this is already partially achieved in the literature. As it turns out, for $R=0$, the theory becomes the $G/G$ gauged WZW-matter model that was introduced in \cite{Okuda:2013fea}. Here, we generalize the approach of \cite{Okuda:2013fea} to formulate an entire family of such theories with a general value of $R$. We shall refer to this new TQFT as the ``equivariant $G/G$ model''.

The fields of the equivariant $G/G$ model with general $R$ are $(A, \lambda, \varphi, \psi, \eta, \chi, g)$, where $A,\varphi,\eta, g$ are bosons and the rest are fermions. The BRST charge $Q_{(g,t)}$ acts on the fields in the following way:
\bea
Q_{(g,t)} A=\lambda, & Q_{(g,t)} \lambda^{(1,0)}=\left(A^g\right)^{(1,0)}-A^{(1,0)}, & Q_{(g,t)} \lambda^{(0,1)}=-\left(A^{g^{-1}}\right)^{(0,1)}+A^{(0,1)}, \nonumber \\
Q_{(g,t)} \varphi=\psi, & Q_{(g,t)} \psi=t\left(\varphi^g\right)-\varphi, & Q_{(g,t)} \psi^{\dagger}=-t\left(\varphi^{\dagger}\right)^{g^{-1}}+\varphi^\dagger, \label{BRST} \\
Q_{(g,t)} \chi=\eta, & Q_{(g,t)} \eta = t \chi^g-\chi, & Q_{(g,t)} \eta^\dagger=-t\left(\chi^{\dagger}\right)^{g^{-1}}+\chi^\dagger, \nonumber \\
Q_{(g,t)} g=0 \,, \nonumber
\eea
where
\bea
A^g & = & g^{-1}A g+g^{-1}d g, \nonumber \\
\varphi^g & = & g^{-1}\varphi g,\\
\chi^g  & = & g^{-1}\chi g \,. \nonumber
\eea
The action of $Q_{(g,t)}$ in (\ref{BRST}) is almost exactly the same as in \cite{Okuda:2013fea}, except that spins of fields (\ref{Spin}) depend on $R$. Also, notice that our conventions here slightly differ from \cite{Okuda:2013fea} by $\eta\leftrightarrow \eta^\dagger$ and $\chi\leftrightarrow \chi^\dagger$.

The square of the BRST charge $Q_{(g,t)}^2=\CL_{(g,t)}$ defines a bosonic transformation on the space of fields and the action of the theory needs to be invariant under it. In the gauged WZW-matter model, the action consists of the original action of the gauged WZW model and a $Q_{(g,t)}$-exact term,
\beq\label{GWZWM}
S_{\mathrm{GWZWM}}=S_{\mathrm{GWZW}}+Q_{(g,t)} (S'),
\eeq
and the theory does not depend on $S'$ as long as the latter satisfies
\beq
\CL_{(g,t)}S'=0.
\eeq
The freedom of choosing different forms of $S'$ can be used to localize the partition function. In the equivariant $G/G$ model with general $R$, the action also takes the form (\ref{GWZWM}):
\beq\label{EGWZW}
S_{R-\mathrm{EGWZW}}=S_{\mathrm{GWZW}}+Q_{(g,t)} (S'),
\eeq
with $S'$ obeying
\beq
\CL_{(g,t)}S'=0.
\eeq
There are different ways to explain why the BRST transformation and the Lagrangian take this particular form. For example, one can start with the Lagrangian and BRST transformation of the $\beta$-deformed complex Chern-Simons theory and compactify on a circle to directly derive the equivariant $G/G$ model. Or, one can start with the UV theory \eqref{SQCD} in 2d and analyze the IR limit following \cite{Witten:1993xi}. Here we will follow a simplified version of the latter approach to illustrate that (\ref{BRST}) and (\ref{EGWZW}) --- which may seem a little strange at a first glance --- are, in fact, what one should expect.

The Lagrangian of the UV theory \eqref{SQCD} consists of two parts. The first part is $\CN=(2,2)$ $U(N)$ SQCD with $k$ fundamental chiral multiplets, which in the IR flows to the gauged WZW model. In the IR, the field $g$ is identified with the scalar component $\sigma$ of the vector multiplet:
\beq
g\sim \sigma.
\eeq
In analyzing the low-energy fate of the second term, we can assume $g=1$. Then, only the mass term remains, and we have
\beq
S_{R-\mathrm{EGWZW}}(A,\lambda,\varphi,\psi,\eta,\chi, g=1)=k S_{\mathrm{GWZW}}(A,\lambda,g=1)+ \int d^2z  \left(m^2\varphi \varphi^\dagger + m\psi \psi^\dagger\right).
\eeq
Indeed, the above action is invariant under $Q_{(1,t)}$ and the second term can be written as
\beq
\int d^2z \left(m^2\varphi \varphi^\dagger+ m\psi \psi^\dagger\right)=Q_{(1,t)}S'=\int d^2z \left[\frac{m}{2} \,Q_{(1,t)} \left(\varphi\psi^\dagger-\psi\varphi^\dagger\right)\right]
\eeq
if we set the IR mass to be
\beq
m=1-t.
\eeq
It is easy to verify that
\beq
\CL_{(1,t)}S'=0.
\eeq
This simplified situation with $g=1$ tells us that the form of the BRST-transformation (\ref{BRST}), which has no derivative terms, and the form of the action (\ref{EGWZW}), where the extra fields only enter via BRST exact terms, are indeed expected.

Now we proceed to find the partition function of the equivariant $G/G$ model with $G=U(N)$ and general $R$. As one would expect, this theory shares a lot of similarities with the gauged WZW-matter model that corresponds to $R=0$ and the localization computation is very similar, except that the spin assignments of various fields can be different. So, instead of repeating everything in section 3 of \cite{Okuda:2013fea}, we only sketch the computation and point out how these two theories are different. First we modify $S'$ to be symmetric in the two BRST-multiplets $(\varphi,\psi)$ and $(\chi,\eta)$ (\textit{cf.} equation (3.15) and (3.16) in \cite{Okuda:2013fea})\footnote{We believe there should be no factor of $k$ multiplying $S_{\mathrm{matter}}$ as appears in \cite{Okuda:2013fea}.}:
\bea
S_{\mathrm{matter}}(g,A,\varphi,\psi, \eta,\chi)&=&Q_{(g,t)}S' \nonumber \\
&=&Q_{(g,t)} \left[\frac{1}{4\pi}\int_\Sigma \Tr\left(\varphi\psi^\dagger-\psi\varphi^\dagger+ \chi\eta^\dagger-\eta\chi^\dagger \right)\right] \\
&=&\frac{1}{2\pi}\int_\Sigma \left\{\left(\varphi-t\varphi^{g},\varphi\right)+\left(\psi,\psi\right)+\left(\chi-t\chi^{g},\chi\right)+\left(\eta,\eta\right)\right\} \,. \nonumber
\eea
Here $(\cdot,\cdot)$ stands for the inner product and its definition for each field is clear from the context.

Now, following \cite{Blau:1993tv}, we perform the abelianization and integrate out the off-diagonal components of $g$, $A$ and $\lambda$. After abelianization, $g$ belongs to the Cartan torus, generated by $H^a,\, a=1,2,\ldots, N$:
\beq
g=\exp\left(2\pi i\sum_{a=1}^N\sigma_a H^a\right),
\eeq
and the fields $(A,\lambda,g)$ are replaced by the abelian fields $(A_a, \lambda_a, \sigma_a)$. Notice that the principal $U(1)^N$-bundle may be non-trivial; it is characterized by the flux $(n_1,\ldots,n_N)$,
\beq
n_a=\frac{1}{2\pi}\int_\Sigma F_a,
\eeq
and we need to sum over all flux sectors. The theory after abelianization is a $BF$-model with $B$ valued in the Cartan torus, coupled to the rest of the fields $(\varphi,\psi,\chi,\eta)$. As all these matter fields have Gaussian action, they can be integrated out explicitly. We first decompose them into the Cartan-Weyl basis that diagonalizes the adjoint action of $g=e^{2\pi i \sigma}$:
\bea
\varphi&=&\sum_{a=1}^N \varphi_a H^a + \sum_{\alpha} \varphi_\alpha E^\alpha, \\
\chi&=&\sum_{a=1}^N \chi_a H^a + \sum_{\alpha} \chi_\alpha E^\alpha, 
\eea
where the $\alpha$'s are the roots of $\frak{su}(N)$ and
\beq
\mathrm{Ad}_{e^{2\pi i \sigma}} (E^\alpha) = e^{2\pi i \alpha(\sigma)} E^\alpha.
\eeq
Upon this decomposition, the trivial adjoint $\frak{u}(N)$ bundle now splits into a direct sum of line bundles $\C^N\oplus \bigoplus_\alpha V_\alpha$ and the fields $\varphi_\alpha$ and $\chi_\alpha$ take values in
\bea
\varphi_\alpha &\in& \Gamma\left[\Omega^0(\Sigma,K^{R/2}\otimes V_\alpha)\right],  \\
\chi_\alpha    &\in& \Gamma\left[\Omega^0(\Sigma,K^{R/2-1}\otimes V_\alpha)\right] \,. 
\eea

Integrating out matter fields valued in the Cartan gives a functional determinant
\beq
\Xi^{\text{2d}}_{\text{ab}}=\prod_{a}^{N}\frac{\mathrm{Det}_\chi (1-t)}{\mathrm{Det}_\varphi (1-t)},
\eeq
while integrating out the matter fields valued in the $V_\alpha$'s will leave us with another functional determinant:
\beq
\Xi^{\text{2d}}_{\text{nab}}=\prod_{\alpha>0}\frac{\mathrm{Det}_\chi \left[M_\alpha(\sigma,t)\cdot M_{-\alpha}(\sigma, t) \right]}{\mathrm{Det}_\varphi \left[M_\alpha(\sigma,t)\cdot M_{-\alpha}(\sigma, t) \right]},
\eeq
where, as in section \ref{sec:3dlocalization},
\beq
M_\alpha(\sigma,t)=1-te^{2\pi i \alpha(\sigma)}.
\eeq
Since $\chi$ is fermionic, the functional determinant associated to it appears in the numerator, while the bosonic determinant for the fields $\varphi$ appears in the denominator. Up to this point, everything is independent of the R-charge assignment of the chiral multiplet $\Phi$ and, in fact, all dependence on the choice of $R$ is encoded in this functional determinant.

As $\chi_\alpha$ and $\varphi_\alpha$ both contain two degrees of freedom, the numerator and the denominator almost cancel. They don't cancel completely because the number of zero modes is different for these two fields. This difference can be computed using the Hirzebruch-Riemann-Roch theorem:
\begin{multline}
\mathrm{dim}\, \Omega^0(\Sigma,K^{R/2-1}\otimes V_{\alpha})-\mathrm{dim}\, \Omega^0(\Sigma,K^{R/2}\otimes V_{\alpha})\\
=1-h+(1-R/2)(2h-2)-\alpha(n)=-\chi(\Sigma)\cdot\frac{1-R}{2}-\alpha(n). \nonumber
\end{multline}
Here $h$ is the genus, $\chi(\Sigma)=2-2h$, and the last term $\alpha(n)$ is the degree of the line bundle $V_\alpha$, which can be written as an integral
\beq
\alpha(n)=\frac{1}{2\pi}\int_\Sigma \alpha(F)=\frac{1}{2\pi}\int_\Sigma \alpha_a F^a.
\eeq

As a result, the first functional determinant is simply
\beq
\Xi_{\text{ab}}(R)=\prod_{i=1}^N(1-t)^{-\chi(\Sigma)\frac{1-R}{2}}=(1-t)^{N(h-1)(1-R)},
\eeq
and the second functional determinant becomes
\begin{multline}
\Xi_{\text{nab}}(R)=\prod_{\alpha>0}\frac{\mathrm{Det}_{(1,0)} \left[M_\alpha(\sigma,t)\cdot M_{-\alpha}(\sigma, t) \right]}{\mathrm{Det}_0 \left[M_\alpha(\sigma,t)\cdot M_{-\alpha}(\sigma, t) \right]}=\prod_{\alpha} M_\alpha(\sigma,t)^{(h-1)(1-R)-\alpha(n)}
\\
=\prod_{\alpha}\left[M_\alpha(\sigma,t)\right]^{(h-1)(1-R)}\cdot e^{-\frac{1}{2\pi}\int_\Sigma \alpha(F)\log M_\alpha} \,. \nonumber
\end{multline}

The partition function for general $R$ is now (\textit{cf.} (3.29) in \cite{Okuda:2013fea})
\begin{multline}
Z^R=\left[\Sigma;U(N),k,t\right]=\frac{1}{|W|}\int\CD\sigma_a\CD \lambda_a\CD A_a \left[\prod_\alpha\left(1-e^{2\pi i (\sigma_a-\sigma_b)}\right)^{1-h}\right]\Xi_{\text{ab}} \Xi_{\text{nab}}\, \\
\times  \exp\left\{i\int_\Sigma\left[ \left((k+N) \sigma_a-\sum_{b=1}^N\sigma_b+\frac{N-1}{2}\right) F^a+\frac{k}{4\pi}\lambda_a \^ \lambda_a\right]\right\}.
\end{multline}
The $F^a$-dependent part of $\Xi_{\text{nab}}$ combines with other terms in the exponent that are proportional to $F^a$ to give
\beq
\zeta_a(\sigma)=k\sigma_a-\frac{i}{2\pi}\sum_{b\neq a}\log \left(\frac{e^{2\pi i \sigma_a}-te^{2\pi i \sigma_b}}{te^{2\pi i \sigma_a}-e^{2\pi i \sigma_b}}\right).
\eeq
Integrating over $A_a$ and over non-zero modes of $\lambda_a$ gives a collection of delta-functions requiring $\zeta_a$ to be integral:
\beq
\sum_{l_a\in \Z}\delta(\zeta_a-l_a).
\eeq
Then we integrate over the $\sigma_a$'s. The delta-functions will produce a factor of
\beq
\sum_{\{\sigma\}\in\{\mathrm{Bethe}\}}\det\left| \frac{\partial \zeta_a}{\partial \sigma_b}\right|^{-1}.
\eeq
Here $\{\mathrm{Bethe}\}$ stands for the set of solutions to the following Bethe ansatz equations:
\beq\label{Bethe2}
e^{2\pi i k\sigma_a}\prod_{b\neq a}\left(\frac{e^{2\pi i \sigma_a}-te^{2\pi i \sigma_b}}{te^{2\pi i \sigma_a}-e^{2\pi i \sigma_b}}\right)=1,\quad \text{for all of $a=1,2,\ldots,N$}.
\eeq
The set of solutions to the Bethe ansatz equations is acted upon by the Weyl group, and after the quotient by this symmetry, the solutions are labeled by Young tableaux with at most $N$ rows and $k$ columns. Notice that the Bethe ansatz equations are the same for all choices of R-charge assignment.

Further integrating over the zero modes of $\lambda_a$ gives a factor
\beq
\left|\frac{\partial \zeta_a}{\partial \sigma_b}\right|^{h}.
\eeq
Therefore, the partition function is
\beq
\boxed{\phantom{\oint}
Z^R(\Sigma;U(N),k,t)=\sum_{\{\sigma\}\in\{\mathrm{Bethe}\}}\left[(1-t)^{N(1-R)}\det\left| \frac{\partial \zeta_a}{\partial \sigma_b}\right|\prod_{a\neq b}\frac{\left(e^{2\pi i \sigma_a}-te^{2\pi i \sigma_b}\right)^{1-R}}{e^{2\pi i \sigma_a}-e^{2\pi i \sigma_b}}\right]^{h-1} \nonumber\phantom{\oint}}
\eeq
This is the partition function of the equivariant $G/G$ model with a general R-charge assignment. Now we proceed to discuss two important cases $R=2$ and $R=0$.

\subsubsection{$R=2$ and the equivariant Verlinde formula}
\label{sec:Rtwo}

As we emphasized earlier, the brane constructions in section \ref{Brane} naturally lead to $R=2$, which is the case that we are mostly interested in. The corresponding 2d TQFT is the equivariant $G/G$ model whose partition function gives the equivariant Verlinde formula:
\beq
Z_{\mathrm{EGWZW}}(\Sigma;U(N),k,t)=\sum_{\{\sigma\}\in\{\mathrm{Bethe}\}}\left[\frac{1}{(1-t)^N }\det\left| \frac{\partial \zeta_a}{\partial \sigma_b}\right|\prod_{a\neq b}\frac{1}{\left(e^{2\pi i \sigma_a}-te^{2\pi i \sigma_b}\right)\left(e^{2\pi i \sigma_a}-e^{2\pi i \sigma_b}\right)}\right]^{h-1}. \nonumber
\eeq
It has several nice properties:
\begin{itemize}
	\item For $t=0$ ($\beta\rightarrow +\infty$), the ``equivariant Verlinde formula'' turns into the ordinary Verlinde formula, as one can directly verify.
	\item In the limit $t\rightarrow 1$ ($\beta\rightarrow 0$), the equivariant Verlinde formula diverges as
	\beq
	Z \sim (1-t)^{-(h-1) \cdot \dim (G)}.
	\eeq
This is indeed what one would expect from the geometry of the Hitchin moduli space, that (up to higher codimension strata) looks like $T^*\CM_{\mathrm{flat}}$. Notice, the order of the pole in the above formula, $(h-1) \cdot \dim (G)$, is precisely the complex dimension of the cotangent fiber, whose non-compactness causes the divergence of the equivariant integral \eqref{ComplexCS2} in the limit $t\rightarrow 1$.
	\item The equivariant Verlinde formula should be a power series with integer coefficients, because it is defined as the graded dimension of the Hilbert space of complex Chern-Simons theory, {\it cf.} \eqref{grdimH} and \eqref{CSPartition}. This is indeed the case, as we will explicitly verify for $G=SU(2)$ in section \ref{EVA}, where a cutting and gluing approach is developed to calculate the same partition function from basic building blocks that only involve rational functions of $t$ that can be written as power series with integer coefficients.
	\item In the limit $k\rightarrow +\infty$, with $k\cdot\beta$ fixed, the equivariant Verlinde formula turns into the formula for the equivariant volume of $\CM_H$, or equivalently, the partition function of the topological Yang-Mills-Higgs model in \cite{Moore:1997dj}.
\end{itemize}

To the best of our knowledge, the equivariant Verlinde formula associated with the choice $R=2$ is novel. In \cite{Gerasimov:2006zt} and \cite{Gerasimov:2007ap}, a model named ``generalized $G/G$ gauged WZW model'' was proposed. Although it shares some similarities with the equivariant $G/G$ model, the BRST-transformation rules, the Bethe ansatz equations and the partition function are all different. It would be interesting to see what the geometric interpretation of the generalized $G/G$ model is, as well as to study its embedding into critical string theory as we have done in section \ref{Brane}.

For the other special value of $R=0$ we get the $G/G$ gauged WZW-matter model of Okuda and Yoshida, which did appeared in the mathematical literature, albeit in a completely different form (as we explain next).

\subsubsection{$R=0$ and gauged WZW-matter model}
\label{sec:Rzero}

For $R=2$, the field $\chi$ is a scalar and $\varphi$ is a 1-form. When $R=0$, their spin assignments are reversed, {\it cf.} (\ref{Spin}).
Therefore, the $h$-dependent parts of the functional determinants are simply inverted when one goes from one case to the other:
\bea
\Xi_{\text{ab}}(R=2)&=&\frac{1}{\Xi_{\text{ab}}(R=0)}=\frac{1}{(1-t)^{N(h-1)}},\\
\Xi'_{\text{nab}}(R=2)&=&\frac{1}{\Xi'_{\text{nab}}(R=0)}=\left[\prod_{\alpha}M_\alpha(\sigma,t)\right]^{1-h}.
\eea
Here
\beq
\Xi'_{\text{nab}}(R)=\left[\prod_{\alpha}M_\alpha(\sigma,t)\right]^{(h-1)(R-1)}
\eeq
is the part of $\Xi_{\text{nab}}$ that does not depend on $F_a$.
So, the partition function of the $G/G$ gauged WZW-matter model is
\beq
Z_{\mathrm{GWZWM}}(\Sigma;U(N),k,t)=\sum_{\{\sigma\}\in\{\mathrm{Bethe}\}}\left[(1-t)^N \det\left| \frac{\partial \zeta_a}{\partial \sigma_b}\right|\prod_{a\neq b}\frac{e^{2\pi i \sigma_a}-te^{2\pi i \sigma_b}}{e^{2\pi i \sigma_a}-e^{2\pi i \sigma_b}}\right]^{h-1}.
\eeq

It was verified numerically in \cite{Okuda:2013fea} that, for small values of $k$, $N$ and $h$, the $G/G$ gauged WZW-matter model gives a 2d TQFT whose corresponding Frobenius algebra is the ``deformed Verlinde algebra'' constructed by Korff in \cite{2013CMaPh.318..173K}. Korff's construction is motivated by the q-boson model and uses the cylindric generalization of skew Macdonald functions.

In fact, the partition function of the gauged WZW-matter model appeared in the mathematical literature even earlier! It can be identified with an index formula for the moduli stack of algebraic $G_\C$-bundles over $\Sigma$ first conjectured by Teleman \cite{2003math......6347T} and later proved by Teleman and Woodward \cite{2003math.....12154T}. As we mentioned earlier, considering the index associated to the prequantum line bundle $\CL$ over $\mathrm{Bun}_{G_\C}(\Sigma)$ --- which is basically $\CM_{\mathrm{flat}}(\Sigma;G)$ away from stacky points --- gives the Verlinde formula. Telemann and Woodward then considered higher rank bundles over $\mathrm{Bun}_{G_\C}(\Sigma)$. In particular, they considered the following bundle:
\beq
\lambda_t(T\CM)\otimes\CL^{\otimes k} \in K^0(\CM,\Q)[t],
\eeq
where $\lambda_t$ stands for the total lambda class, defined as follows. For a vector bundle $V$ over space $X$, let $\lambda^l(V)$ be the $K^0$-class of $\Lambda^l V$, then
\beq
\lambda_t(V)=1+t\lambda^1(V)+t^2\lambda^2(V)+...\in K^0(X,\Q)[t].
\eeq
One can explicitly check that, at least for $G=U(N)$, the index of this bundle can be identified with the partition function of the gauged WZW-matter model, modulo a sign convention for the equivariant parameter:
\beq
t_{\mathrm{TW}} \; = \; -t_{\mathrm{here}} \,.
\eeq

\subsection{Relation to Bethe/Gauge correspondence}
\label{sec:BetheGauge}

In \cite{Nekrasov:2009uh,Nekrasov:2009ui,Nekrasov:2009rc}, Nekrasov and Shatashvili proposed a relation between integrable models and supersymmetric gauge theories with four supercharges. In this paper, we are concerned with two types of 3d $\CN=2$ theories: $T[L(k,1);\beta]$ and $T[\Sigma\times S^1;\beta]$. Of course, these two theories are special cases of $T[M_3;\beta]$, where $M_3$ is an arbitrary Seifert manifold.

The theory $T[\Sigma\times S^1;\beta]$, which was the subject of section~\ref{SUSYSide}, does not have a Chern-Simons term; it is a canonical mass deformation of 3d $\CN=4$ theory. The relation between theories of this type and integrable models was also explored in \cite{Gadde:2013wq,Gaiotto:2013bwa}.
Here, we shall focus on the Lens space theory $T[L(k,1);\beta]$.

Although Okuda and Yoshida \cite{Okuda:2013fea} found a relation between the gauged WZW-matter model and the q-boson model, the connection to SUSY gauge theory was missing. The results of our work fill this gap. In particular, according to our discussion in section \ref{ComplexTwist}, the gauged WZW-matter model is precisely 3d $\CN=2$ theory $T[L(k,1);\beta]$ twisted on $\Sigma\times S^1$. This kind of scenario was discussed by Nekrasov and Shatashvili in \cite{Nekrasov:2014xaa}, and we now embed $T[L(k,1);\beta]$ into the framework of Bethe/gauge correspondence following their work.

{}From the matter content \eqref{thetwothys} of $T[L(k,1);\beta]$, one can easily write down the effective twisted superpotential:
\beq\label{Weff}
\tilde{\CW}_{\mathrm{eff}}(\sigma)=(k+N)\pi i\sum_{a=1}^N \sigma_a^2-\pi i\left(\sum_{a=1}^N\sigma_a\right)^2+\frac{1}{2\pi i}\sum_{a \neq b}\Li_2\left[te^{2\pi i(\sigma_a-\sigma_b)}\right]+(N-1)\pi i \sum_{a=1}^N\sigma_a.
\eeq
The first two terms come from the Chern-Simons term of the 3d $\CN=2$ vector multiplet. One can directly see that, after integrating out W-bosons, the levels for the $SU(N)$ and the $U(1)$ parts of the $U(N)$ gauge group are now $k+N$ and $k$, respectively. The third term in \eqref{Weff} comes from the adjoint chiral multiplet in 3d. And, unsurprisingly,
\beq
t=e^{-\ell \beta},
\eeq
where $\ell$ is the circumference of the $S^1$ Seifert fiber. The last term in \eqref{Weff} also originates from 3d gauge fields:
\be
2\pi i\langle \rho,\sigma\rangle=\pi i\sum_{a>b}(\sigma_a-\sigma_b)\sim \pi i (N-1)\sum_{a=1}^N\sigma_a.
\ee
Here $\rho$ is the Weyl vector, and, in the last step, we have used the fact that the shift
\be
\tilde{\CW}_{\mathrm{eff}}\longrightarrow \tilde{\CW}_{\mathrm{eff}}+2\pi i \sum_a^N n_a\sigma_a
\ee
generates a symmetry of the 2d abelian system.

The Bethe ansatz equations are given by
\beq
\exp\left[\frac{\partial \tilde{W}_{\mathrm{eff}}}{\partial \sigma_a}\right]=e^{2\pi i\zeta_a}=1,\quad \text{ for all $a=1,2,\dots, N$.}
\eeq
The topological action is
\beq
\int_\Sigma \left[\frac{\partial \tilde{W}_{\mathrm{eff}}}{\partial \sigma_a}F_a+\frac{1}{4\pi i}\frac{\partial^2 \tilde{W}_{\mathrm{eff}}}{\partial \sigma_a \partial \sigma_b}\lambda^a\^\lambda^b + \CU(\sigma)\CR\right],
\eeq
where the last term involves the Euler density $\CR$ and the dilaton coupling
\beq
\CU(\sigma)=\CU_\mathrm{gauge}(\sigma)+\CU_\mathrm{matter}(\sigma),
\eeq
such that
\beq
\CU_\mathrm{gauge}(\sigma)=\sum_{\alpha}\log\left(1-e^\alpha(\sigma) \right),
\eeq
and
\beq
\CU_\mathrm{matter}(\sigma)=\left(\frac{R-1}{2}\right)\Tr_{\mathrm{adj}}\left[\log\left(1-te^{-\sigma}\right)\right].
\eeq
Here and throughout the paper,
$R$ is the $U(1)_V$ R-charge assigned to $\Phi$ and, in fact, this is the only place where $R$ enters our formula.
For the two choices of R-charge assignment discussed in sections \ref{sec:Rtwo} and \ref{sec:Rzero} we have:
\beq
\CU(\sigma)_\mathrm{matter}^{R=0} \; = \; -\,\CU(\sigma)_\mathrm{matter}^{R=2} \,.
\eeq
Then, the partition function of the topologically twisted theory is written as a sum over solutions to the Bethe ansatz equations:
\beq
Z^{R}\left(T[L(k,1);\beta];\Sigma\times S^1\right)=\sum_{\{\sigma\}\in\{\mathrm{Bethe}\}} \left(e^{-\CU^R(\sigma)}\det\left|\frac{1}{2\pi i}\frac{\partial^2 \tilde{W}_{\mathrm{eff}}}{\partial \sigma_a\partial\sigma_b}\right| \right)^{g-1}.
\eeq
One can check that this expression indeed agrees with the partition functions obtained previously. In particular, it gives the equivariant Verlinde formula for $R=2$ and the partition function of the gauged WZW-matter model for $R=0$.

Each summand in the partition function of the twisted SUSY gauge theory should be mapped to the squared norm of a Bethe state on the integrable model side. (Bethe states have a natural normalization and their norms are physical quantities.) As was checked in \cite{Okuda:2013fea}, the summands in the partition function of the gauged WZW-matter model indeed correspond to the squared norms of Bethe states of the q-boson model. Naturally, this raises a series of questions: What about the partition functions of topological theories with $R\neq 0$? What is their meaning on the integrable model side? Is $R=0$ ``special''?

It would be also interesting to study (quantum) spectral curves for 3d $\CN=2$ theories $T[L(k,1);\beta]$ and $T[M_3;\beta]$ following \cite[sec.~5]{Gadde:2013wq}. The spectral curves for these theories are expected to be spectral curves of integrable systems related to the ones discussed here by spectral duality. In particular, it should provide a candidate for the spectral duality of the q-boson model, and it would be interesting to make contact with~\cite{Mironov:2013xva}.


\section{$t$-deformation and categorification of the Verlinde algebra}
\label{EVA}

In previous sections, we focused on the partition function of the $\beta$-deformed complex Chern-Simons theory on $\Sigma\times S^1$ --- the equivariant Verlinde formula --- and have shown that it can be derived in at least three different ways (the first is intrinsically three-dimensional and the other two are two-dimensional):
\begin{enumerate}

\item Section \ref{sec:3dlocalization}: Starting with the 3d $\CN=2$ theory $T[L(k,1);\beta]$ one can perform a topological twist on $\Sigma\times S^1$ and compute the partition function using localization {\it \`a la} \cite{Ohta:2012ev}.

\item Section \ref{sec:equivariantGG}: One can first reduce twisted $T[L(k,1);\beta]$ to 2d to obtain the equivariant $G/G$ gauged WZW model on $\Sigma$ and apply localization techniques and compute its partition function as in \cite{Okuda:2013fea}.

\item Section \ref{sec:BetheGauge}: One can first compactify $T[L(k,1);\beta]$ on a circle and obtain the low-energy effective $\CN=(2,2)$ abelian gauge theory governed by the twisted effective superpotential as a function on the Coulomb branch. Then one can twist this 2d theory and compute its partition function following \cite{Nekrasov:2014xaa}.

\end{enumerate}

Naturally, the next step is to go beyond the partition function and incorporate operators. Indeed, one would expect loop operators to play very interesting role in complex Chern-Simons theory, just as they do in ordinary Chern-Simons theory. Recall that in Chern-Simons theory with compact gauge group $G$, Wilson loops are labeled by integrable representations of the loop group $LG$ and their fusion rules give the Verlinde algebra, which basically describes how the tensor product of two representations decomposes. Then one can ask what the analog of this story in the $\beta$-deformed complex Chern-Simons theory is.

It turns out that a finite $\beta$ simply deforms the Verlinde algebra to what we call the ``equivariant Verlinde algebra''. For example, the usual fusion rule for $G=SU(2)$ at level $k=9$ for two fundamental representations
\beq
\mathbf{2}\otimes\mathbf{2}=\mathbf{1}\oplus \mathbf{3}
\label{22usual}
\eeq
is deformed into
\beq
\mathbf{2}\otimes\mathbf{2}=\frac{1}{1-t^2}\mathbf{1}\oplus \frac{1}{1-t}\mathbf{3}\oplus\frac{t}{1-t}\mathbf{5}\oplus\frac{t^2}{1-t}\mathbf{7}\oplus\frac{t^3}{1-t^2}\mathbf{9} \,.
\label{22deformed}
\eeq
Clearly, in the limit $t \to 0$ ($\beta \to \infty$) one recovers \eqref{22usual}. Expressions like \eqref{22deformed} are ubiquitous in computations of refined BPS invariants and categorification of quantum group invariants \cite{Gukov:2011ry,Fuji:2012pi}. In fact, just like in those examples, each coefficient on the right-hand side of \eqref{22deformed} is a graded dimension \eqref{grdimH} of an infinite-dimensional vector space $V_j$ that appears as a ``coefficient'' in the OPE of line operators in 3d:
\beq
\mathbf{2}\otimes\mathbf{2}= \bigoplus_j V_j \otimes (\mathbf{2j+1}).
\label{22categor}
\eeq
In other words, as explained {\it e.g.}~in \cite{Gukov:2006jk,Gukov:2007ck}, replacing $\Sigma\times S^1$ by $\Sigma\times \R$ leads to a categorification in the sense that numerical coefficients are replaced by vector spaces (whose dimensions are the numerical coefficients). In the present case, we obtain a categorification of the equivariant Verlinde algebra since the ``coefficients'' in the OPE of line operators on $\Sigma\times \R$ are indeed vector space, namely $V_j$ in our case. In the present example, \eqref{22categor} is a categorification of \eqref{22deformed} with
\bea
V_0 & = & \C [x_0] \{ 0 \}, \nonumber \\
V_1 & = & \C [x_1] \{ 0 \}, \nonumber \\
V_2 & = & \C [x_2] \{ 1 \}, \\
V_3 & = & \C [x_3] \{ 2 \}, \nonumber \\
V_4 & = & \C [x_4] \{ 3 \}, \nonumber
\eea
where $\dim_{\beta} (x_0) = \dim_{\beta} (x_4) = 2$, $\dim_{\beta} (x_1) = \dim_{\beta} (x_2) = \dim_{\beta} (x_3) = 1$, and $\{ n \}$ denotes the degree shift by $n$ units.

Of course, one can obtain a deformed algebra such as \eqref{22deformed} by computing the partition function on $\Sigma\times S^1$ with insertion of multiple loop operators that lie along the $S^1$ fiber direction, using similar localization techniques as in previous sections. This problem will be studied more systematically elsewhere. In this section, we will analyze a simplified version of this problem with $G=SU(2)$ using a completely different method. Namely, we evaluate the equivariant integrals over Hitchin moduli space directly for some simple Riemann surfaces and build the TQFT using cutting and gluing. 

\subsection{``Equivariant Higgs vertex''}

In order to perform cutting and gluing, it is important to generalize everything to punctured Riemann surfaces. We use $\Sigma_{h,n}$ to denote a Riemann surface with genus $h$ and $n$ ramification points $p_1,p_2,\ldots,p_n \in \Sigma$. Here we only consider ``tame'' ramification discussed in detail in \cite{Gukov:2006jk,Gukov:2007ck}. Near each puncture $p_r$, the ramification data is specified by a triple denoted as\footnote{This triple is denoted as $(\alpha,\beta,\gamma)$ in \cite{Gukov:2006jk,Gukov:2007ck}. Here we use a different notation to avoid confusion with the equivariant parameter $\beta$.} $(\alpha_I,\alpha_J,\alpha_K)\in \mathbf{T}^3$, where $\mathbf{T}=U(1)$ is the Cartan torus of $G=SU(2)$. However, our approach only applies directly to cases where $\alpha_J=\alpha_K=0$, as $U(1)_{\beta}$, which we use to regularize the non-compactness of the moduli space, acts on $\alpha_J+i\alpha_K$ by multiplying it with a phase. In order to make it invariant under $U(1)_{\beta}$, we need to impose the condition $\alpha_J=\alpha_K=0$. In the following, we simply use $\alpha_r$ to denote $\alpha_I$ associated with the ramification point $p_r$.

Then the moduli space of ramified Higgs bundles $\CM_H(\Sigma_{h,n};\alpha_1,\alpha_2,\ldots,\alpha_n)$ can be identified with the moduli space of flat $SL(2,\C)$ connections over $\Sigma_{h,n}$ with boundary condition that near a puncture $p_r$, only the real part $A$ of the connection $\CA=A+i\phi$ develops a singularity
\beq
A\sim \alpha_r d\theta.
\eeq
Equivalently, we demand the holonomy around each puncture $p_r$ to be in the same conjugacy class as
\beq
e^{2\pi i\alpha_r\sigma^3}=\exp\left[2\pi i\left(\begin{array}{cc}
\alpha_r&0\\
0&-\alpha_r
\end{array}\right)\right].
\eeq
The action of the affine Weyl group on $\alpha$'s leaves the conjugacy class of the monodromy invariant. So without loss of generality, we assume all $\alpha_r$'s to live in the Weyl alcove $[0,\frac{1}{2}]$.

As in the unramified case, we can consider the problem of quantizing $\CM_H(\Sigma_{h,n};\alpha_1,\alpha_2,\ldots,\alpha_n)$ with symplectic form $k\omega_I$ and our goal is to identify a 2d TQFT whose partition function is the dimension of the Hilbert space $\CH(\Sigma_{h,n};\alpha_1,\alpha_2,\ldots,\alpha_n)$. This TQFT --- which we call $SU(2)$ ``equivariant Verlinde TQFT'' --- is equivalent to the equivariant G/G model of section \ref{sec:Rtwo} specialized to the choice of $G=SU(2)$, but formulated in a different way, via cutting and gluing.

Any 2d TQFT can be formulated in a set of Atiyah-Segal axioms, which assign a Hilbert space $V$ to a circle $S^1$ and an element in $\mathrm{Hom}(V^{\otimes n},\C)$ to a punctured Riemann surface $\Sigma_{h,n}$. In particular, if $n=0$, the TQFT assigns to a genus-$h$ Riemann surface an element in $\mathrm{Hom}(\C,\C)$. This element is determined by the image of $1\in\C$, which is precisely the partition function in physicists' language.

Two-dimensional TQFTs are particularly simple, as any Riemann surface, punctured or not, can be cut along circles to be decomposed into three basic ingredients: the cap, the cylinder and pair of pants, {\it cf.}~figure \ref{fig:Genus2}. One only needs to determine how the TQFT functor acts on the three basic building blocks. If we find a basis $e_\mu$ (or in physicists' notation $\{\langle \mu|\}$) of $V$, then the TQFT assigns ``metric'' $\eta^{\mu\nu}$ to a cylinder, ``fusion coefficients'' $f^{\mu\nu\rho}$ to a pair of pants, and a distinguished state $e_{\O}\in V$ to a cap. This is summarized in table~\ref{TQFTRules}.
\begin{table}
\begin{center}
\begin{tabular}{ccc}
	\raisebox{-0.45\totalheight}{\includegraphics[width=1.8cm]{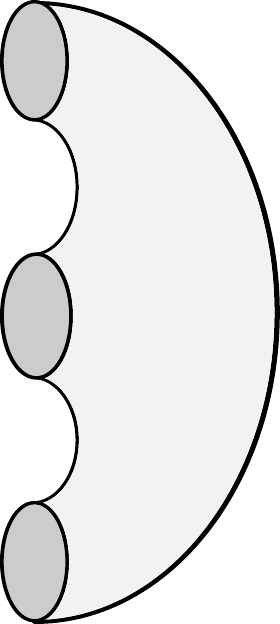}} &$\leadsto$&$f^{\mu\nu\rho}$\\
	\\
	\raisebox{-0.45\totalheight}{\includegraphics[width=1.8cm]{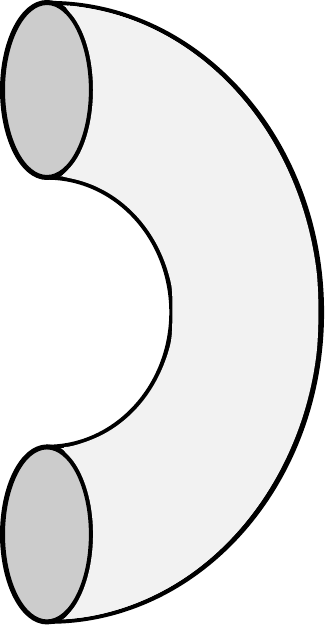}}&$\leadsto$&$\eta^{\mu\nu}$\\
	\\
	\raisebox{-0.5\totalheight}{\includegraphics[width=0.8cm]{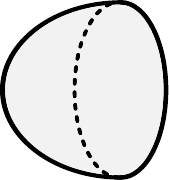}}     &$\leadsto$ & $e_{\O}$
\end{tabular}
\end{center}
\caption{\label{TQFTRules} Building blocks of a 2d TQFT.}
\end{table}

Topological invariance requires the ``equivariant Higgs vertex'' $f^{\mu\nu\rho}$ to be symmetric in the three indices. Also, as a four-holes sphere can be decomposed into two pairs of pants in different ways, the fusion coefficients have to satisfy the commutativity relation:
\beq\label{Commutativity}
f^{\mu_1\nu_1\rho_1}\eta_{\rho_1\rho_2}f^{\mu_2\nu_2\rho_2}=f^{\mu_1\nu_2\rho_1}\eta_{\rho_1\rho_2}f^{\mu_2\nu_1\rho_2}.
\eeq
Here $\eta_{\rho_1\rho_2}=\left(\eta^{-1}\right)^{\rho_1\rho_2}$ is the inverse metric naturally defined on $V^{*\otimes 2}$. Using these properties, it is easy to prove that a 2d TQFT is equivalent to a commutative Frobenius algebra. For the equivariant Verlinde TQFT, the corresponding algebra is the ``equivariant Verlinde algebra'', the one parameter generalization of the Verlinde algebra that we alluded to.

Before figuring out what $V,\eta^{\mu\nu},f^{\mu\nu\rho}$ and $e_{\O}$ are, we first see what the prediction from the equivariant gauged WZW model looks like. First of all, the dimension of $V$ should be the number of solutions to the Bethe ansatz equations
\beq
\mathrm{dim}\, V= Z_{\mathrm{EGWZW}}\left[\T^2;SU(2)\right]=\sum_{\{\mathrm{Bethe}\}} 1.
\eeq
The Bethe ansatz equations for $SU(2)$ can be obtained\footnote{There are two ways of eliminating the $U(1)$ factor. Apart from the one described here, one can also set $\sigma_1=-\sigma_2$. This corresponds to $U(2)/U(1)=SU(2)/\Z_2=SO(3)$.} by combining the two equations for $U(2)$,
\bea
e^{2\pi i k\sigma_1}\left(\frac{e^{2\pi i\sigma_1}-te^{2\pi i\sigma_2}}{te^{2\pi i\sigma_1}-e^{2\pi i\sigma_2}}\right)=1,\\
e^{2\pi i k\sigma_2}\left(\frac{e^{2\pi i\sigma_2}-te^{2\pi i\sigma_1}}{te^{2\pi i\sigma_2}-e^{2\pi i\sigma_1}}\right)=1,
\eea
into a single equation satisfied by
\beq
\sigma=\frac{1}{2}(\sigma_1-\sigma_2)\in \left[0,\frac{1}{2}\right].
\eeq
So the Bethe ansatz equation for $SU(2)$ is simply
\beq\label{BetheSU2}
e^{4\pi i k\sigma}\left(\frac{e^{2\pi i\sigma}-te^{-2\pi i\sigma}}{te^{2\pi i\sigma}-e^{-2\pi i\sigma}}\right)^2=1.
\eeq

In the limit $\beta\rightarrow +\infty$ ($t\rightarrow 0$), the equivariant Verlinde TQFT becomes the ordinary Verlinde TQFT ({\it i.e.} $G/G$ WZW model) and the Bethe ansatz equation becomes:
\beq\label{Bethe0}
e^{4\pi i (k+2)\sigma}=1.
\eeq
There are $k+1$ solutions to this equation, namely:
\beq
\sigma_l=\frac{l+1}{2(k+2)}, \quad l=0,1,\ldots,k \,.
\eeq
One can verify that this number of solutions is independent of $\beta$ and will always be $k+1$. So, regardless of the value of $\beta$, the Hilbert space $V$ of a 2d TQFT is always $k+1$-dimensional.

There is one subtle point that is worth mentioning. In the literature there is some confusion about the ``end point contribution'' to the Verlinde formula. Namely, $l=-1$ and $l=k+1$ also give valid solutions to the equation (\ref{Bethe0}) and they indeed appear in localization computation (see {\it e.g.} \cite{Blau:1993tv}). However, their contribution is divergent if genus $h>1$, and it was argued that they should be simply ignored. Our approach gives a different point of view on this issue. For any positive value of $\beta$, solutions $\sigma_{-1}$ and $\sigma_{k+1}$ are never inside the interval $[0,\frac{1}{2}]$ and, therefore, they never contribute to the equivariant Verlinde formula. When $\beta\rightarrow +\infty$, we have $\sigma_{-1}\rightarrow 0$ from the left and $\sigma_{k+1}\rightarrow \frac{1}{2}$ from the right. If we think of the ordinary Verlinde formula as the $\beta\rightarrow \infty$ limit of the equivariant Verlinde formula, then we should never include the contributions associated to $\sigma_{-1}$ and $\sigma_{k+1}$. Similar phenomena happen when $\beta\rightarrow 0$. In that limit, $\sigma_0$ and $\sigma_k$ move toward the endpoints of $[0,\frac{1}{2}]$. But as they will always be inside the interval, one should always include their full contributions.

The fact that $V$ is finite dimensional is also expected from the geometry of the Hitchin moduli space. If there is no puncture, then $\CM_H(\Sigma_{h,n};SU(2))$ with symplectic form $k\omega_I$ is always quantizable. However, if we add punctures, $k\omega_I$ may not have integral periods over all 2-cycles of $\CM_H(\Sigma_{h,n};\alpha_1,\alpha_2,\ldots,\alpha_n)$, and this will be an obstruction to quantization. So, the $\alpha$'s need to satisfy certain integrality conditions that we now analyze.

In general, the moduli space of a ramified Higgs bundle can be conveniently viewed as a fibration of coadjoint orbits over the moduli space of unramified Higgs bundles. More concretely, in the case of $G_\C=SL(2,\C)$, we have
\beq
\begin{matrix}T^*\bbCP^1_{\alpha_1}\times \ldots \times T^*\bbCP^1_{\alpha_n} & \to & M_H(\Sigma_{h,n};\alpha_1,\alpha_2,\ldots,\alpha_n)\\
                                &   & \downarrow \\
                                & & M_H(\Sigma_{h,n}),
\end{matrix}
\eeq
where $T^*\bbCP^1_{\alpha_r}=\CO_{\alpha_r}$ is the orbit of $\alpha_r$ in $\frak{sl}(2,\C)$ under adjoint action. Then integrality of the periods of $k\omega_I$ is translated to the following condition:
\beq
\int_{\bbCP^1_{\alpha_r}}k\omega_I=2k\alpha_r\in \Z.
\eeq
If we introduce
\beq
\lambda_r=2k \alpha_r \in[0,k],
\eeq
then there are $k+1$ possible values of $\lambda_r$ for each puncture $p_r$, corresponding to $k+1$ states $\langle\lambda|$'s in $V$. And this indeed agrees with the prediction of the equivariant gauged WZW model. These states correspond to point-like defects on the Riemann surface, and from the three-dimensional point of view, these defects are Wilson loops along the $S^1$ fiber direction of $\Sigma\times S^1$.

Another prediction from physics is that the partition function of the equivariant Verlinde TQFT --- or, equivalently, the value of the equivariant integral \eqref{ComplexCS2} over $\CM_H$ --- can be naturally written as a sum over solutions to the Bethe ansatz equation. A similar phenomenon was already pointed out back in \cite{Moore:1997dj}, but it was never verified or properly understood. Next, we will construct the TQFT and see how the Bethe ansatz equation for $SU(2)$ naturally arises when one attempts to diagonalize the fusion rules.

\subsection{Equivariant Verlinde algebra from Hitchin moduli space}

In order to derive the ``equivariant Higgs vertex'' $f^{\lambda_1\lambda_2\lambda_3}$, we do the equivariant integration over the Hitchin moduli space, $\CM_H(\Sigma_{0,3};\alpha_1,\alpha_2,\alpha_3)$, associated with the three-punctured sphere. The virtual dimension of this space is $2\times(3h-3+n)=0$, so we expect it to be a collection of points which makes the equivariant integration very easy. We first consider the limit $\beta\rightarrow +\infty$. In this limit, the equivariant integral becomes an ordinary integral over the moduli space of $SU(2)$ connections and simply counts the number of points in $\CM = \CM_{\mathrm{flat}}(\Sigma_{0,3}; SU(2), \alpha_1,\alpha_2,\alpha_3)$. In fact, this moduli space is either a point or empty. So the fusion coefficient $f^{\lambda_1\lambda_2\lambda_3}_{\beta\rightarrow +\infty}$ is either 1 or zero. One special thing about this zero-dimensional moduli space is that the quantizability condition is slightly more subtle, as the coadjoint orbits $\bbCP^1_{\alpha_i}$'s are no longer real 2-cycles. More precisely, in addition to requiring $(\alpha_1,\alpha_2,\alpha_3)$ to satisfy integrality condition
\beq
(\lambda_1,\lambda_2,\lambda_3)=2k(\alpha_1,\alpha_2,\alpha_3)\in\Z^3,
\eeq
one also needs to require $\lambda_1+\lambda_2+\lambda_3$ to be even.
Then, the condition for $f^{\lambda_1\lambda_2\lambda_3}_{\beta\rightarrow+\infty}$ to be 1 is that $(\lambda_1,\lambda_2,\lambda_3)$ satisfies both the quantization condition and the ``triangle inequality''. We now explain the second condition more precisely, which is important for the equivariant generalization later.

\begin{figure}[htb]
	\centering
	\includegraphics[width=0.70\textwidth]{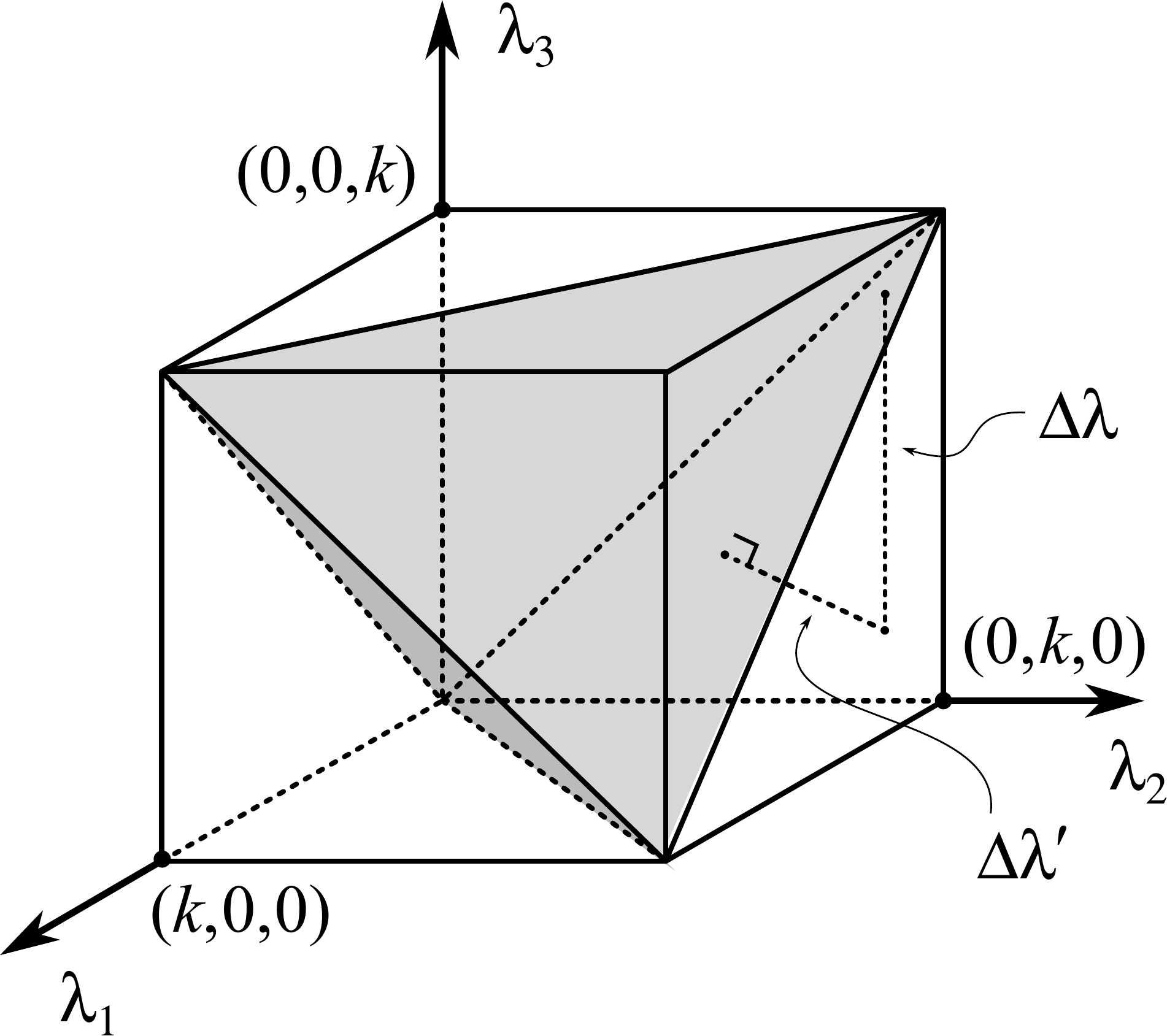}
	\caption{\label{fig:Tetrehedron} The definition of $\Delta\lambda$ and $\Delta\lambda'=\Delta\lambda/\sqrt{3}$.}
\end{figure}

When the quantization condition is satisfied, the triple $(\lambda_1,\lambda_2,\lambda_3)$ corresponds to an integer point in the cube $\{(x,y,z)|0\leq x,y,z\leq k\}$. There is a tetrahedron inside this cube with four faces given by the following four equations:
\bea\nonumber
d_0&=&\lambda_1+\lambda_2+\lambda_3-2k=0,\\\nonumber
d_1&=&\lambda_1-\lambda_2-\lambda_3=0,\\\nonumber
d_2&=&\lambda_2-\lambda_3-\lambda_1=0,\\
d_3&=&\lambda_3-\lambda_1-\lambda_2=0.
\eea
Define the distance of a point $(\lambda_1,\lambda_2,\lambda_3)$ to the tetrahedron faces as, see figure \ref{fig:Tetrehedron},
\beq
\Delta\lambda=\mathrm{max}(d_0,d_1,d_2,d_3).
\eeq
We also define another quantity
\beq
\Delta\alpha=\frac{\Delta\lambda}{2 k}.
\eeq
If $\Delta\lambda\leq 0$, then the point is either inside the tetrahedron or on the boundary of it and $\CM_{\mathrm{flat}}(\Sigma_{0,3}; SU(2))$ is a point. If $\Delta\lambda>0$, then the point is outside the tetrahedron and $\CM_{\mathrm{flat}}(\Sigma_{0,3}; SU(2))$ is empty. We call this condition ``triangle inequality'' for the following reason: when $d_1>0$ or $d_2>0$ or $d_3>0$, the three $\lambda$'s won't be able to form a triangle. The situation $d_0>0$ corresponds to the case when the triangle is too large to live in $SU(2)$, which is a compact group.

Combining the quantization condition with the $\Delta\lambda\leq 0$ condition, we obtain the fusion coefficient in the $\beta\rightarrow +\infty$ limit:
\beq
f^{\lambda_1\lambda_2\lambda_3}=\left\{\begin{array}{cl} 1 & \textrm{if $\lambda_1+\lambda_2+\lambda_3$ is even and $\Delta\lambda\leq 0$,}\\
0 & \textrm{otherwise.}\end{array}\right.
\eeq

We now consider the case of finite $\beta$. The geometry of the relevant Hitchin moduli space $\CM_H$ is described in detail in \cite{1996alg.geom.10014B}. What differs from the $\beta\rightarrow +\infty$ case is that $\CM_H(\Sigma_{0,3};\alpha_1,\alpha_2,\alpha_3)$ is never empty and is always a point. This is consistent with a general property of moduli space of parabolic Higgs bundles: the topology only depends on the quasi-parabolic structure. Then, the equivariant integral
\beq
\int_{\CM_H}\mathrm{Td}(\CM_H,\beta)\^e^{k\tilde{\omega}_I}
\eeq
simply becomes
\beq
e^{-\beta k\mu_0}=e^{-\beta\Delta\lambda/2}=t^{\Delta\lambda/2},
\eeq
where
\beq
\mu_0=\Delta\alpha
\eeq
is the value of the moment map for $U(1)_{\beta}$ at that point \cite{1996alg.geom.10014B}. So we have the fusion coefficients
\beq\label{Fusion}
f^{\lambda_1\lambda_2\lambda_3}=\left\{\begin{array}{ll} 1 & \textrm{if $\lambda_1+\lambda_2+\lambda_3$ is even and $\Delta\lambda\leq 0$,}\\
e^{-\beta\Delta\lambda/2} & \textrm{if $\lambda_1+\lambda_2+\lambda_3$ is even and $\Delta\lambda> 0$,}\\
0 & \textrm{if $\lambda_1+\lambda_2+\lambda_3$ is odd.}\end{array}\right.
\eeq

The next thing one needs is the metric $\eta^{\mu\nu}$ associated to a cylinder. As the Hitchin moduli space $\CM_H\left(\Sigma_{0,2};SU(2)\right)$ has negative virtual dimension, one needs to be careful when trying to make sense of the equivariant integral. Alternatively, one can deduce $\eta^{\mu\nu}$ from the equivariant Verlinde number associated with other Riemann surfaces. For example, one can consider the four-holed sphere and do the integration over $\CM_H\left(\Sigma_{0,4};SU(2)\right)$. This moduli space is an elliptic surface with the elliptic fibration over $\C$, which is precisely the Hitchin fibration. The only singular fiber is the ``nilpotent cone'', the fiber over zero of the Hitchin base $\C$, and has Kodaira type $I_0^*$ (or, affine $D_4$ in physicists' notation), see {\it e.g.} \cite{Gukov:2007ck} for details. These nice properties make the equivariant integration easy to do. But instead of presenting the results of this computation, we directly give the form of the metric that is obtained by combining this result with the fusion coefficients:
\beq\label{Metric}
\eta^{\lambda_1\lambda_2}=\mathrm{diag}\{1-t^2,\underbrace{1-t,\,\,1-t,\,\,\ldots,\,\,1-t,}_\text{$k-1$ entries that are all $(1-t)$}1-t^2\}.
\eeq
Notice that because $\CM_H\left(\Sigma_{0,2};SU(2)\right)$ has virtual complex dimension $-2$, $\eta$ has a first order zero when $t\rightarrow 1$, instead of a pole. Also, the metric is diagonal and only becomes the identity matrix when $t=0$.

{}Once we know $f$ and $\eta$, it is easy to find the state $\langle \O |$ from the consistency of the gluing rules (attaching a cap to a pair of pants should give a cylinder):
\beq
f^{\mu\nu\O}=\eta^{\mu\nu}.
\eeq
And one finds
\beq
\langle \O |=\langle 0|-t\langle 2|,
\eeq
when $k\geq 2$. For $k=1$ and $k=0$, $\langle \O |=\langle 0|$ and one can further verify that the Verlinde TQFT is not deformed by turning on $\beta$ in these two cases.

Before proceeding further it is convenient to introduce a normalized basis $\{\langle \underline{\lambda}|=\left(\eta_{\lambda\lambda}\right)^{1/2}\langle \lambda|\}$ in which the TQFT ``metric'' $\eta$ is the identity. In this basis, the commutativity relation (\ref{Commutativity}) becomes simply
\beq
f^{\lambda_1\underline{\mu\nu}}f^{\lambda_2\underline{\nu\xi}} \; = \; f^{\lambda_2\underline{\mu\nu}}f^{\lambda_1\underline{\nu\xi}} \,.
\eeq
The above relation can be interpreted as the mutual commutativity of $k+1$ matrices $[f^0]$, $[f^1]$, $\ldots$, $[f^{k+1}]$, where
\beq
[f^\mu]^{\underline{\nu\rho}} = f^{\mu\underline{\nu\rho}},
\eeq
is a $(k+1)\times(k+1)$ matrix.

Now, that we have all the building blocks of the equivariant Verlinde TQFT, we can calculate any correlation function on any Riemann surface. However, the basis $\{\langle 0|$, $\langle 1|$, $\ldots$ $\langle k|\}$, or its normalized version, is not the most convenient for this purpose. One would like to work in a different basis $\{\langle \hat{0}|$, $\langle \hat{1}|$, $\ldots$, $\langle \hat{k}|\}$ where the fusion rules are diagonalized. Namely, in the normalized basis, the matrices $[f^0]$, $[f^1]$, $\ldots$, $[f^{k}]$ are mutually commutative and simultaneously diagonalizable, with $\{\langle \hat{0}|$, $\langle \hat{1}|$, $\ldots$, $\langle \hat{k}|\}$ being the set of eigenvectors. As the fusion coefficients $f^{\mu\nu\rho}$ are completely symmetric in the three indices, in the diagonal basis we have
\beq
f^{\hat{\mu}\hat{\nu}\hat{\rho}} \sim \delta_{\hat{\mu}\hat{\nu}\hat{\rho}},
\eeq
where $\delta_{abc}$ is the ``3d Kronecker delta function'' (equal to 1 when $a=b=c$ and zero otherwise).

Before attempting to find this new basis, we first briefly comment on its normalization. There are two possible choices: we can choose either
\beq
f^{\hat{\mu}\hat{\nu}\hat{\rho}}=\delta_{\hat{\mu}\hat{\nu}\hat{\rho}},
\eeq
or
\beq
\eta^{\hat{\mu}\hat{\nu}}=\delta^{\hat{\mu}}_{\hat{\nu}}.
\eeq
If the first normalization is chosen, this basis is what mathematicians would call the ``idempotent basis'' of the equivariant Verlinde algebra and it coincides with the basis formed by ``Bethe states''. We will work with the second choice of normalization, where one does not need to distinguish between upper and lower indices.

\subsection{Bethe Ansatz equation from the fusion rules}

The standard way to find the eigenvectors of a set of commuting matrices is to first pick a linear combination of the matrices and to solve for the eigenvalues. At this point, one may (correctly) anticipate that the characteristic polynomial equation of a particular linear combination of $[f]$'s gives the Bethe ansatz equation. Indeed, this is true and that matrix is
\beq
[f_B]=[f_1]-t[f_3],
\eeq
when $k\geq 3$. For $k=2$, $[f_B]=[f_1]$ does not depend on $\beta$ at all. As it turns out, $\beta$ only appears in the normalization factor when $k=2$, making this case uninteresting. So from now on, we assume that $k\geq 3$. Written in the matrix form, $f_B$ is
\beq
[f_B]=\left(\begin{array}{ccccccc}
0   & \sqrt{1+t} & 0 & 0 & 0 & \cdots & 0\\
\sqrt{1+t} & 0   & 1 & 0 & 0 &\cdots & 0\\
0   & 1   & 0 & 1 &  \phantom{\sqrt{1+t}}  &    \phantom{\sqrt{1+t}}   &\vdots\\
0   & 0   & 1 &\ddots&\ddots& & 0\\
0   & 0   & &\ddots&0 &1      & 0\\
\vdots & \vdots& \phantom{\sqrt{1+t}} &\phantom{\sqrt{1+t}}  & 1 & 0 & \sqrt{1+t}\\
0 & 0 &\cdots &0  &0   &\sqrt{1+t}   &0
\end{array}\right).
\eeq
The characteristic polynomial equation for $[f_B]$ is
\beq\label{CharEq}
\det\left(x [I]-[f_B]\right)=0,
\eeq
where $[I]$ is the identity matrix of size $(k+1)\times (k+1)$. By expanding this determinant along the first and last columns, it is easy to find that
\beq
\det\big(x [I]-[f_B]\big)=x^2 A_{k-1}-2x (1+t) A_{k-2} + (1+t)^2 A_{k-3},
\eeq
where $A_n$ is a polynomial in $x$ defined as the determinant of a $n\times n$ matrix
\beq
A_n=\det\left(\begin{array}{ccccccc}
x   & -1 & 0 & 0 & 0 & \cdots & 0\\
-1 & x   & -1 & 0 & 0 &\cdots & 0\\
0   & -1   & x & -1 &   &       &\vdots\\
0   & 0   & -1 &\ddots&\ddots& & 0\\
0   & 0   & &\ddots&x &-1      & 0\\
\vdots & \vdots& & & -1 & x & -1\\
0 & 0 &\cdots &0  &0   &-1   &x
\end{array}\right).
\eeq
Using the initial condition $A_0=1$ and $A_1=x$, along with the recursion relation
\beq
A_{n+1}=x A_{n}- A_{n-1}
\eeq
that can be derived by expanding the determinant along the first column, one finds
\beq
A_n=\frac{\sin\left[2\pi (n+1)\sigma\right]}{\sin 2\pi\sigma} \,.
\eeq
Here we made the following change of variables
\beq
x \; = \; 2\cos2\pi\sigma \,.
\eeq
Then, one finds the characteristic polynomial equation (\ref{CharEq}) to be
\beq\label{Bethe3}
e^{4\pi ik\sigma}\left(\frac{e^{2\pi i\sigma}-te^{-2\pi i\sigma}}{te^{2\pi i\sigma}-e^{-2\pi i\sigma}}\right)^2=1 \,.
\eeq
This is exactly the Bethe ansatz equation (\ref{BetheSU2}) for the equivariant $SU(2)/SU(2)$ gauged WZW model!

For $0< t< 1$, the equation (\ref{Bethe3}) always has $k+1$ real solutions $\sigma_l$, $l=0,1,\ldots,k$ inside the interval $\left(0,\frac{1}{2}\right)$. So we can assume $\sigma_0<\sigma_1<...\sigma_{k}$. As we mentioned previously, in the limit $t\rightarrow 0$, the Bethe ansatz equation (\ref{Bethe3}) becomes
\beq
e^{4\pi i (k+2) \sigma} \; = \; 1 \,.
\eeq
And in the other limit $t\rightarrow 1$, the Bethe ansatz equation (\ref{Bethe3}) becomes
\beq
e^{4\pi i k \sigma} \; = \; 1 \,.
\eeq
This agrees with the fact that the quantum shift of the level $k$ in Chern-Simons theory with complex gauge group is zero \cite{Gukov:2008ve}.

There is another interesting property satisfied by the Bethe ansatz equation (\ref{Bethe3}). If $\sigma \in(0,\frac{1}{2})$ is a solution to (\ref{Bethe3}), then $\frac{1}{2}-\sigma$ is also a solution. So the $k+1$ roots $\{\sigma_l\}$ are naturally paired. As a consequence, if $k$ is even, $\sigma_{k/2}=\frac{1}{4}$ is always a solution.

Now we have the eigenvalues $x_{l}=2\cos2\pi\sigma_{l}$ that are solutions to (\ref{Bethe3}), and the next step is to find the eigenvectors $\langle\hat{l}|$. In the normalized basis, they are $(k+1)\times 1$ matrices $[v_l]$ that can be obtained by solving the linear equation
\beq
[f_B][v_{l}]=x_{l}[v_{l}].
\eeq
It is easy to find
\beq\label{Eigenvector}
[v_{l}]=C_l\left(
\begin{array}{c}
\sqrt{1+t}\sin2\pi\sigma_{l}\\
\sin4\pi\sigma_{l}\\
\sin6\pi\sigma_{l}-t\sin2\pi\sigma_{l}\\
\sin8\pi\sigma_{l}-t\sin4\pi\sigma_{l}\\
\vdots\\
\sin 2k\pi\sigma_{l}-t\sin 2(k-2)\pi\sigma_{l}\\
\frac{\sin2(k+1)\pi\sigma_{l}-t\sin 2(k-1)\pi\sigma_{l}}{\sqrt{1+t}}
\end{array}
\right).
\eeq
Here $C_l$ is a normalization factor
\beq
C_l^{-2}=\left|1-te^{4\pi i\sigma_{l}}\right|^2\cdot\frac{k+2}{2}+2t\cos 4\pi\sigma_{l}-2t^2 \,.
\eeq
In the new basis, the fusion rules are:
\beq
f^{\hat{\mu}\hat{\nu}\hat{\rho}}=N_{\hat{\mu}}\delta_{\hat{\mu}\hat{\nu}\hat{\rho}} \,.
\eeq
Explicitly, the ``eigenvalues of the fusion rules'' $N_l$'s are
\beq
N_{l}=\frac{1}{\sqrt{1-t}\cdot\sin 2\pi\sigma_{l}\left|1-te^{4i\pi\sigma_{l}}\right|^2} \,.
\eeq
In particular, from gluing $2h-2$ copies of pairs of pants (as in figure~\ref{fig:Genus2}), it immediately follows
that on a closed Riemann surface $\Sigma_h$ the partition function is
\bea
Z(\Sigma_h;k,t) & = & \sum_{l=0}^k N_l^{2h-2} \label{EqVerSU2} \\
& = & \frac{1}{(1-t)^{h-1}}\sum_{l=0}^k\left(\frac{k+2}{2}+\frac{2t\cos4\pi\sigma_l-2t^2}{\left|1-te^{4\pi i\sigma_l}\right|^2}\right)^{h-1}\left(\frac{1}{\sin2\pi\sigma_l\left|1-te^{4\pi i\sigma_{l}}\right|}\right)^{2h-2}. \nonumber
\eea
We call this the ``$SL(2,\C)$ equivariant Verlinde formula''. It is easy to check that for $t=0$ it indeed reduces to the $SU(2)$ Verlinde formula. In the special case of $h=0$, we have
\beq
Z(S^2;k,t)=\sum_{\hat{l}=0}^k N_{\hat{l}}^{-2}=\sum_{\hat{l}=0}^k|\langle\underline{\hat{l}}|\phi\rangle|^2 =\langle\phi|\phi\rangle= 1-t^3.
\eeq
For generic values of $t$, this formula gives a non-trivial identity satisfied by roots of the Bethe Ansatz equation.

To a $n$-punctured Riemann surface, the 2d TQFT functor assigns a vector in $\left(V^*\right)^{\otimes n}$:
\beq
Z(\Sigma_{h,n};k,t)=\sum_{l=0}^k N_{l}^{2h-2+n}\langle\hat{l}|^{\otimes n}.
\eeq


\acknowledgments{We wish to thank Anton Alekseev for a wonderful set of notes \cite{Alekseev:2000fe} that we recommend to all the beginners.
We also thank S.~Shatashvili for discussions of this work in Fall 2013 and Spring 2014, which stimulated \cite{Nekrasov:2014xaa}.
We also benefited from discussions with Mina Aganagic, Sir Michael Atiyah, Tudor Dimofte, Abhijit Gadde, Jaume Gomis, Nigel Hitchin, Tadashi Okazaki, Satoshi Okuda, Pavel Putrov, Richard Wentworth and Wenbin Yan.

This work is funded by the DOE Grant DE-SC0011632, NSF grants DMS 1107452, 1107263, 1107367 (the GEAR Network), and the Walter Burke Institute for Theoretical Physics.}


\newpage

\bibliographystyle{JHEP_TD}
\bibliography{equivariant}

\end{document}